 \documentclass[12pt,preprint]{aastex}

\newcommand{\mjb}{mJy~beam$^{-1}$}

\newcommand\kms{km~s$^{-1}$}

\shorttitle{H$_2$CO Masers}
\shortauthors{Hoffman et al.}

\begin{document}

\title{The Formaldehyde Masers in NGC~7538 and G29.96-0.02: VLBA, MERLIN, and VLA Observations}

\author{Ian M.\ Hoffman}
\affil{National Radio Astronomy Observatory, P.\ O.\ Box O, Socorro, NM, USA 87801}
\affil{Department of Physics and Astronomy, University of New Mexico, Albuquerque, NM, USA 87131}
\email{ihoffman@nrao.edu}

\author{W.\ M.\ Goss}
\affil{National Radio Astronomy Observatory, P.\ O.\ Box O, Socorro, NM, USA 87801}

\author{Patrick Palmer}
\affil{Department of Astronomy and Astrophysics, 5640 S.\ Ellis Avenue, Chicago, IL, USA 60637}

\and

\author{A.\ M.\ S.\ Richards}
\affil{MERLIN/VLBI National Facility, University of Manchester, Jodrell Bank Observatory, Macclesfield, Cheshire SK11 9DL, UK}

\begin{abstract}
The 6~cm formaldehyde (H$_2$CO) maser sources in the compact \ion{H}{2} regions NGC~7538-IRS1 and G29.96-0.02 have been imaged at high resolution ($\theta_{beam} < 50$~mas).
Using the VLBA and MERLIN, we find the angular sizes of the NGC~7538 masers to be $\sim 10$~mas (30~AU) corresponding to brightness temperatures $\sim 10^8$~K.
The angular sizes of the G29.96-0.02 masers are $\sim 20$~mas (130~AU) corresponding to brightness temperatures $\sim 10^7$~K.
Using the VLA, we detect 2~cm formaldehyde absorption from the maser regions.
We detect no emission in the 2~cm line, indicating the lack of a 2~cm maser and placing limits on the 6~cm excitation process.
We find that both NGC~7538 maser components show an increase in intensity on 5-10 year timescales while the G29.96-0.02 masers show no variability over 2 years.
A search for polarization provides 3-$\sigma$ upper limits of 1\% circularly polarized and 10\% linearly polarized emission in NGC~7538 and of 15\% circularly polarized emission in G29.96-0.02.
A pronounced velocity gradient of 28~${\rm km}\,{\rm s}^{-1}\,{\rm arcsecond}^{-1}$ (1900~${\rm km}\,{\rm s}^{-1}\,{\rm pc}^{-1}$) is detected in the NGC~7538 maser gas.
\end{abstract}


\section{Introduction}

Astronomical observation of emission from the $1_{10} \rightarrow 1_{11}$ transition at 4.83~GHz (6~cm) of the formaldehyde (H$_2$CO) molecule is exceedingly rare.
Only four Galactic sources have been observed to emit: three of these sources have been shown to be masers.
The H$_2$CO 6~cm emission in the compact \ion{H}{2} region NGC~7538~B was first observed by Downes \& Wilson (1974) and was shown to be a compact, non-thermal maser by Forster et al.\ (1980).
Most recently, an H$_2$CO maser was discovered in G29.96-0.02 by Pratap, Menten \& Snyder (1994, hereafter PMS94).
H$_2$CO masers are also observed in Sgr~B2 (Whiteoak \& Gardner 1983).
The fourth H$_2$CO emission source, Orion-KL, is known to have extended thermal emission ({\it e.g}.\ Johnston et al.\ 1983).
Though detection of H$_2$CO masers is expected to require interferometric observations which are not sensitive to the broad angular scale $1_{10} \rightarrow 1_{11}$ absorption that dominates single dish measurements of the 4.83~GHz line ({\it e.g}.\ Bieging et al.\ 1980), dedicated interferometric surveys ({\it e.g}.\ Forster et al.\ 1985; PMS94; Mehringer, Goss, \& Palmer 1995) have been rather unsuccessful in discovering new masers.

There is currently no working theoretical model of the Galactic formaldehyde maser process, despite a history of relatively rapid understanding of astrophysical formaldehyde phenomena.
Almost immediately after the discovery of interstellar formaldehyde in 6~cm absorption by Snyder, Buhl, Zuckerman, \& Palmer (1969), observations were made of ``anomalous'' 6~cm absorption (Palmer, Zuckerman, Buhl, \& Snyder 1969), 2~cm absorption (Evans, Cheung, \& Sloanaker 1970), and emission in the $2_{12} \rightarrow 1_{11}$ (Kutner, Thaddeus, et al.\ 1971) and $2_{11} \rightarrow 1_{10}$ (Thaddeus et al.\ 1971) millimeter lines (see Fig.~1).
Improved laboratory measurements (Tucker, Tomasevich, \& Thaddeus 1971, 1972; Nerf 1972) and excitation calculations (Townes \& Cheung 1969; Thaddeus 1972; Evans 1975; Garrison et al.\ 1975) explained these phenomena shortly thereafter, but cannot explain the observed maser emission.
Boland \& de Jong (1981) proposed a continuum radiation pump mechanism to explain the NGC~7538 maser.
Though the radiative pump has been successfully applied to the several ($> 10$) extragalactic H$_2$CO (mega)masers which have been observed ({\it e.g}.\ Baan et al.\ 1986, 1993), the model is not applicable to the continuum morphology discovered in NGC~7538 more recently ({\it e.g}.\ Pratap et al.\ 1992), nor can the model explain the Sgr~B2 and G29.96-0.02 masers discovered subsequently.
Thus, the Galactic 6~cm formaldehyde maser phenomenon remains unexplained nearly 30 years after its discovery.
In a search for new empirical constraints on these mysterious objects, this paper presents an observational study of the Galactic H$_2$CO masers in NGC~7538 and G29.96-0.02.

NGC~7538 (S158) is a well-studied \ion{H}{2} region complex at a distance of $d \approx 3$~kpc (Campbell \& Thompson 1984).
At the southern edge of the optical nebula lie three radio and infrared sources.
The radio sources were designated A, B, \& C by Martin (1973) from 2$\arcsec$-resolution 5~GHz observations.
Wynn-Williams et al.\ (1974) detected infrared counterparts at 20~$\mu$m for the radio sources; A: IRS2, B: IRS1, and C: IRS3.
The masers are located in front of IRS1 (IRAS 23116+6111) which is partially optically thick at 6~cm.
The IRS1 \ion{H}{2} region is thought to have a central star of type O6~V or B0.5~II ({\it e.g}.\ Israel et al.\ 1973; Willner 1976) and  there is a relatively large mass of cold dust associated with IRS1 (but not IRS2 or 3) resulting in a relatively low \ion{H}{2} region gas-to-dust mass ratio of 75 (Willner 1976).
NGC~7538-IRS1 is also associated with a remarkable number of maser species besides H$_2$CO (see \S4.4).

G29.96-0.02 (hereafter G29.96) is a cometary \ion{H}{2} region at a distance of $d \approx 6.5$~kpc (Morisset et al.\ 2002).
The H$_2$CO masers lie in a ``hot core'' about two arcseconds west of the cometary head.
The hot core ($T > 100$~K; $n > 10^7\ {\rm cm}^{-3}$) is thought to be powered by an embedded star, not by the cometary \ion{H}{2} region ({\it e.g}.\ De~Buizer et al.\ 2002; Pratap et al.\ 1999; Cesaroni et al.\ 1998).
The hot core also shows maser emission from H$_2$O and CH$_3$OH (Hofner \& Churchwell 1996; Walsh et al.\ 1998).

All previous observational studies of H$_2$CO masers have utilized either single dish antennas or $\sim 10$~km baseline interferometers ({\it e.g}.\ Forster et al.\ 1980, 1985; Rots et al.\ 1981; PMS94).
Past observations were unable to spatially resolve the maser emission.
A lower limit of $T_B \simeq 10^5$~K brightness temperature has been determined for the NGC~7538 and G29.96 masers (Rots et al.\ 1981; PMS94).
The NGC~7538 H$_2$CO source has two velocity components at $v_{\rm LSR} \simeq -58$~\kms\ (component I) and $-60$~\kms\ (component II), separated by $\sim 0\farcs1$ (Rots et al.\ 1981).
Forster et al.\ (1985) noted a factor of two intensity increase in component~I over 3 years.
No polarization was detected in the NGC~7538 maser components to a 3-$\sigma$ upper limit of 5\% of the total intensity in the Stokes $Q$, $U$, and $V$ parameters (Forster et al.\ 1985).
The G29.96 masers exhibit several blended spectral features including velocity components at $v_{\rm LSR} \simeq 100$~\kms\ (component I) and $102$~\kms\ (component II) that appear coincident on the sky (PMS94).
No polarization observations have been made of the G29.96 masers.

Sensitive milliarcsecond-resolution observations of H$_2$CO masers are needed.
Placing more stringent empirical constraints on Galactic H$_2$CO 6~cm maser emission requires the following observational results:
(1) a determination of the angular extent of the maser emission $\theta_m$, allowing a determination of the intrinsic sizes and brightness temperatures,
(2) a detection of the radiatively coupled $2_{11} \rightarrow 2_{12}$ (2~cm) line, yielding limits on inversion and level populations,
(3) sensitive limits on polarized intensity,
and (4) a characterization of the observed maser intensity increase in NGC~7538.
Furthermore, accurate relative positions of the H$_2$CO masers with respect to other maser species could constrain the poorly-understood physical conditions of the maser environment.

Compared to the Rots et al.\ (1981) VLA observations of the NGC~7538 H$_2$CO masers which used 12 antennas recording a single polarization, the VLA is now more sensitive, using all 27 antennas and recording dual polarizations.
Furthermore, using the sensitivity of the VLBA with the phased VLA, observations of the relatively weak ($\sim 100$~mJy) H$_2$CO masers at milliarcsecond angular resolution are possible.
We have conducted new observations of the NGC~7538 and G29.96 H$_2$CO masers at 6~cm and 2~cm using the Very Large Array (VLA) of the NRAO\footnote{The National Radio Astronomy Observatory (NRAO) is a facility of the National Science Foundation operated under a cooperative agreement by Associated Universities, Inc.} and at 6~cm using Very Long Baseline Array (VLBA) of the NRAO and the Multi-element Radio Linked Interferometry Network\footnote{MERLIN is operated as a National Facility by the University of Manchester, Jodrell Bank Observatory, on behalf of Particle Physics and Astronomy Research Council (PPARC).} (MERLIN).

\section{Observations and Data Reduction}

We have summarized the observational parameters in Tables 1, 2, and 3.
Tables 1 and 2 list the logistical parameters of the observations including bandwidth, spectral configuration, and calibrator information.
All antennas in all observations use dual-circular polarization feeds.
Table 3 lists the resulting sensitivities of the images.
We assume throughout that the observed 6~cm formaldehyde line is the dominant $2 \rightarrow 2$ hyperfine component which is separated from the $1 \rightarrow 0$ component by 1.13~\kms\ (Tucker et al.\ 1971; Downes et al.\ 1976).

The maser line widths, velocities, and velocity gradients were determined using the The Groningen Image Processing System (GIPSY) software package\footnote{\tt http://www.astro.rug.nl/${\mathtt \sim}$gipsy/}.
All other reduction and imaging was done with the Astronomical Image Processing System (AIPS) software package\footnote{\tt http://www.nrao.edu/aips/}.

\subsection{VLA `CnB'}

The NGC~7538 and G29.96 masers were observed on 13 February 1996 at 4.83~GHz (6~cm) using the 27 antennas of the VLA for 20 minutes each.
The NGC~7538 and G29.96 maser positions were also observed on 13 February 1996 at 14.49~GHz (2~cm) with the VLA for 40 minutes each.
In `CnB' configuration the VLA baseline lengths range from 0.065~km to 7.7~km.

All of the VLA `CnB' observations were sensitive to the continuum radiation from the \ion{H}{2} regions.
The continuum contribution was subtracted from the maser channels using the AIPS task UVLSF.
The 2~cm continuum radiation from G29.96 was averaged and imaged from 40 line-free channels over about 1~MHz of bandwidth.
The RMS noise of the line-free 2~cm continuum image of G29.96 is 3.0~\mjb.

\subsection{VLA `B'}

The NGC~7538 masers were observed on 24 July 2002 at 4.83~GHz with the VLA for 40 minutes.
In `B' configuration the array baseline lengths range from 0.21~km to 11.4~km.

The VLA `B' observations were sensitive to the continuum radiation from the \ion{H}{2} regions.
The continuum contribution was subtracted from the maser channels using the AIPS task UVLSF.
The continuum radiation was averaged and imaged from 28 line-free channels over 84~kHz of bandwidth.
The RMS noise of the line-free continuum image is 1.6~\mjb.

\subsection{MERLIN}

The NGC~7538 masers were observed on 1-3 October 2001 using the MERLIN radio telescope of Jodrell Bank Observatory for a total of $\sim 28$ hours.
Six antennas were used; the Mark II at Jodrell Bank, the 32~m antenna at Cambridge, and the 25~m dishes at Knockin, Darnhall, Tabley, and Defford.
The baseline lengths of MERLIN range from 11~km to 217~km.
The phases were calibrated by frequent observations at 14~MHz bandwidth of 2300+638.
3C84 was observed at both 14~MHz and 500~kHz bandwidth allowing bandwidth-dependent calibration to be transferred.

Due to a pointing error, the MERLIN observations were centered 1.6 arcminutes from the maser position.
Though a $2\arcmin$ displacement is within the primary receiving beam of the MERLIN antennas (${\rm FWHM} \sim 10\arcmin$), decoherent averaging during on-line integration (`time smearing') made necessary a factor of 2.8 amplitude correction (see Thompson 1999).
The measured positions of the masers were not affected by the error.

\subsection{VLBA+Y27}

Using the ten antennas of the VLBA and the 27-element phased VLA as an 11-station VLBI array, we observed the NGC~7538 and G29.96 H$_2$CO masers on 22 September 2000 for $\sim 1.8$~hours each.
The amplitude scale was set by online system temperature monitoring and {\it a priori} gain measurements.
The VLBA observations of the masers were phase-referenced to their respective phase calibrators.
The phase referencing technique allows the maser visibility phases (which correspond to sky position) to be determined from those of the brighter phase-reference source (see Walker 1999).
The maser targets and the phase-reference sources were alternately observed in a 2.5:1.5~minute cycle.
Data from the Hancock, North Liberty, Mauna Kea, and St.\ Croix VLBA stations were not part of the final data set because the station-based visibility phases could not be adequately tracked and transferred due to poor signal to noise.
The baseline lengths contributing to the final images range from 52 to 1806~km.

\section{Results}

The VLA images of the continuum radiation from the \ion{H}{2} regions are shown in Figures~2 and 3.
The VLA `B' array image (\S2.2) of the 6~cm continuum radiation from NGC~7538 is shown in Figure~2 with the position of the H$_2$CO masers indicated.
The VLA `CnB' array image (\S2.1) of the 2~cm continuum radiation from G29.96 is shown in Figure~3 with the position of the H$_2$CO masers indicated.

\subsection{NGC~7538 6~cm Emission}

\subsubsection{Angular Position and Size}

We have detected H$_2$CO masers with VLBI observations for the first time.
The VLBA images of the NGC~7538 H$_2$CO component I masers are shown in Figure~4.
Component I splits into two peaks, Ia and Ib, in the VLBA image.
The absolute position uncertainty of the strongest peak in the VLBA images of component I is 2~mas, phase-referenced to the calibrator J2302+6405.
The positions of the emission peaks in the brightest VLBA channel image at $v_{\rm LSR} = -57.84$~\kms\ are summarized in Table~4.
The VLBA position for component I is consistent with Rots et al.\ (1981) within their 100-mas-uncertainty.
The Ia and Ib feature peaks are separated by $14.0 \pm 0.5$~mas (42~AU) in the $v_{\rm LSR} = -57.84$~\kms\ channel image.
The total extent of the velocity structure of component I in the VLBA images is $\sim 60$~mas ($\sim 200$~AU).

The MERLIN images show both components I and II.
Figure~5 shows the channel images of the brightest emission from the two components.
The components are separated by $79 \pm 5$~mas (240~AU), in agreement with the previous determination of $110 \pm 30$~mas by Rots et al.\ (1981).

Component I shows a very compact structure from which MERLIN observations recover all of the emission and from which the VLBA recovers about 60\% of the emission.
In the VLBA images (Fig.~4), features Ia and Ib have resolved angular diameters $\theta_m \approx 10$~mas corresponding to brightness temperatures $T_B \approx 10^8$~K.
In contrast, component II does not show any compact structure on long baselines: MERLIN detects only $\sim 70$\% of the emission and only the two shortest VLBA baselines detect component II.
The deconvolved size of component II in the MERLIN image (Fig.~5) is 64~mas.
The component sizes are consistent with the previous upper limits of 150~mas measured by Rots et al.\ (1981).
The angular size of component II corresponds to a brightness temperature of $5 \times 10^6$~K.

The measurement of the maser brightness temperature and of the background continuum radiation which the maser amplifies allows a determination of the gain of the maser.
The measurements are related by $T_B \approx T_{bg} e^{-\tau} $ where $T_B$ is the brightness temperature of the maser, $T_{bg}$ is the brightness temperature of the background continuum radiation, and $\tau$ is the gain of the maser.
Using a brightness temperature $T_{bg} = 2700$~K measured from the VLA `B' observations, the masers have gains $\tau \simeq -11$.

\subsubsection{Velocity Gradient}

The maser lines are resolved in velocity in all of the observations except those from the VLA `CnB' array.
All of the lines are fit well by single gaussians.
Table~5 summarizes the fitted and deconvolved peak intensity, velocity width at half-maximum, and line center for the different observations.

The MERLIN and VLBA observations show, for the first time, a velocity gradient in the masing gas.
In the VLBA images, the dashed arrow in Figure~4 traces the direction of a velocity field which extends through the Ia peak at a position angle of $31 \pm 5\arcdeg$ for several tens of milliarcseconds (several beamwidths).
Figure~6 shows a position-velocity plot of the slice indicated in Figure~4.
The plot shows a linear velocity dependence along the slice of magnitude 28~${\rm km}\,{\rm s}^{-1}\,{\rm arcsecond}^{-1}$ (1900~${\rm km}\,{\rm s}^{-1}\,{\rm pc}^{-1}$).
The MERLIN observations show a velocity gradient in both components that is consistent with that shown in Figure~6.

We note that a linear velocity gradient is consistent with the maser emission arising from a rigid rotating disk (or cylinder).
However, the position of component II is not consistent with a linear extension of the component I gradient shown Figure~6.
That is, component II does not lie on the disk which models Figure~6.
Emission from a maser component at $v_{\rm LSR} = -60.19$~\kms\ consistent with the 28~${\rm km}\,{\rm s}^{-1}\,{\rm arcsecond}^{-1}$ velocity gradient would lie $\approx 60$~mas away from component I at $\approx 30\arcdeg$ position angle, as marked by the `$+$' symbol in Figure~5.
However, component II is observed to lie $\approx 80$~mas from component I at $17\arcdeg$ position angle (Fig.~5).
Therefore we conclude that the two maser components are not simply related and do not lie in a single, uniform gas cloud.

\subsubsection{Intensity}

The VLA `B' array spectrum of the H$_2$CO masers is shown in Figure~7.
In July 2002, we observed component~I to have an intensity of $1257 \pm 5$~\mjb.
The intensity on July 2002 is a factor of three larger than the $335 \pm 20$~\mjb\ observed in April 1983 by Forster et al.\ (1985) when variability was first noted using the Westerbork Synthesis Radio Telescope with 0.36~\kms\ velocity resolution.
We note an intensity increase in component II for the first time.
In July 2002, we observe an intensity for component II of $385 \pm 5$~\mjb\ with the VLA `B' array compared to $205 \pm 20$~\mjb\ observed by Forster et al.\ (1985).
A light curve of the maser component intensities is shown in Figure~8.

\subsubsection{Polarization}

At the 3-$\sigma$ level, we detect no polarization in any of the observations.
Our VLA data place a more sensitive upper limit of 1\% (3-$\sigma$) on circularly polarized intensity than the previous limit of 5\% set by Forster et al.\ (1985).
The MERLIN and VLBA data show no detectable linearly polarized intensity, corresponding to a 10\% (3-$\sigma$) upper limit in Stokes $Q$ \& $U$, consistent with the previous limit of 5\% set by Forster et al.\ (1985).

\subsection{G29.96 6~cm Emission}

\subsubsection{Angular Position and Size}

Table~6 summarizes the VLBA image properties of the G29.96 masers.
The VLBA+Y27 $v_{\rm LSR} = 100.2$~\kms\ image of G29.96 H$_2$CO maser component I is shown in Figure~9.
The absolute position uncertainty of the VLBA image peak is 10~mas, phase-referenced to the calibrator J1851+0035.
The image of component II in the $v_{\rm LSR} = 102.1$~\kms\ channel appears coincident on the sky.
The two components are separated by $1 \pm 3$~mas ($7 \pm 19$~AU).

The resolved sizes of the masers in the VLBA images are $\theta_m \simeq 20$~mas, corresponding to brightness temperatures $T_B \sim 10^7$~K.
Using a brightness temperature $T_{bg} = 50$~K for the background 6~cm continuum radiation measured from the VLA `CnB' observations after a beam dilution correction using the 2~cm continuum image, the G29.96 maser gains are $\tau \simeq -12$.

\subsubsection{Line Widths}

Figure~10 shows the G29.96 H$_2$CO maser spectra.
The figure includes both the VLBA+Y27 spectrum from the current paper and the VLA `A' array spectrum observed by Pratap et al.\ (PMS94).
Pratap et al., with a spectral resolution of 0.38~\kms, detected ``several blended features'' whose emission is spread over $\sim 5$~\kms.
Remarkably, we detect two {\em spectrally unresolved} velocity features in the VLBA data despite a more favorable spectral resolution of 0.30~\kms.
Our VLA `CnB' spectra are consistent with Pratap et al.\ after correction for absorption, which is discussed below in \S3.3.2.
The properties of the G29.96 line emission are summarized in Table~7.

\subsubsection{Polarization}

At the 3-$\sigma$ level, we detect no polarization in any of the observations.
The VLA and VLBA observations were sensitive only to Stokes $I$ and $V$ parameters.
The VLA data place an upper limit of 15\% (3-$\sigma$) circularly polarized intensity on component I.
The VLBA data are also consistent with zero polarized intensity.

\subsection{Absorption}

\subsubsection{Two Centimeter}

We have detected the 2~cm transition in the direction of Galactic H$_2$CO maser sources for the first time.
Using the VLA `CnB' array, we observe the 2~cm line in absorption in both NGC~7538 and G29.96.
All of the 2~cm line detections have a corresponding 6~cm feature that is coincident in velocity.
The optical depths $\tau$ of all of the absorption features are noted in Figures 11-14.
The optical depths are calculated using
$$ \tau = -\ln\left(1-\frac{T_L}{T_x - T_{bg}}\right)\ ,$$
where $T_L$ is the (negative) temperature of the line (after continuum subtraction) and $T_x \approx 2$~K is the excitation temperature of the line (Palmer et al.\ 1969; Thaddeus 1972).
The derived H$_2$CO column densities are discussed in \S4.

Figure 11 shows the 2~cm absorption at the position of the NGC~7538 masers.
The 2~cm minimum in the NGC~7538 observations corresponds to the position and velocity ($v_{\rm LSR} \simeq -60$~\kms) of (the weaker) maser component II (Fig.~11b).
The position of the 2~cm minimum at the component~I velocity ($v_{\rm LSR} \simeq -58$~\kms, Fig.~11d) lies less than two arcseconds away.

Figure~12 shows the 6~cm and 2~cm spectra at the position of the G29.96 maser.
No 2~cm line is detected at the velocity of the maser but an absorption feature is seen at $v_{\rm LSR} \simeq 107$~\kms.
The 6~cm absorption evident in Figure~12a at $v_{\rm LSR} \simeq 107$~\kms\ and $v_{\rm LSR} \simeq 98$~\kms\ is discussed below.

No 2~cm line emission is detected from either NGC~7538 or G29.96 with a 3-$\sigma$ brightness temperature sensitivity limit of 43~K.
A low gain ($\tau \simeq -1$) maser amplification of the 2~cm doublet in either NGC~7538 or G29.96 would produce an emission line of brightness temperature $T_B \gtrsim 100$~K.
We conclude that there is no inversion of the 2~cm line.

\subsubsection{Six Centimeter}

Using the VLA `CnB' array we detect 6~cm absorption in both NGC~7538 and G29.96.
We detect 6~cm absorption in G29.96 at the position of the masers in the hot core (Fig.~12a) as well as against the cometary \ion{H}{2} region.
Figure~13a shows the fit to the hot core absorption features which are summarized in Table~8.
The maser spectrum after subtraction of the absorption (Fig.~13b) is consistent with the Pratap et al.\ (PMS94) observations.
Figure~14a shows the spectrum at the 6~cm minimum in the VLA `CnB' observations which corresponds to absorption against the cometary \ion{H}{2} region.
Figure~14b shows the 2~cm absorption spectrum at this position.

We detect no 6~cm absorption at the position of the masers in NGC~7538-IRS1.
Figure~14c shows the spectrum at the 6~cm minimum of the VLA `CnB' observations which corresponds to absorption against IRS2.
No 2~cm absorption is detected at this position (Fig.~14d) which is consistent with the 2~cm observations of Mart\'{\i}n-Pintado et al.\ (1985) and Evans et al.\ (1975) (though compare our Fig.~11a,b with Fig.~9 of Mart\'{\i}n-Pintado et al.).

\subsubsection{Column Densities}

The formaldehyde column densities can be calculated from the absorption spectra.
Column densities $N$ calculated from 6~cm spectra are denoted $N_6$ while those calculated from 2~cm spectra are denoted $N_2$.
The following relations were used to determine column densities from the spectra ({\it e.g}.\ Wilson 1972; Wadiak et al.\ 1988; Osterbrock 1989)
$$\frac{N_6}{T_x} = 1.36 \times 10^{13} \sum \tau \Delta{v}\ ,$$
$$\frac{N_2}{T_x} = 1.26 \times 10^{13} \sum \tau \Delta{v}\ ,$$
where $N$ has units ${\rm cm}^{-2}$, $T_x$ has units K, $v$ has units \kms, and where the summation is over the spectral channels with line radiation.
The column densities are summarized in Table~9 grouped by cloud velocity and background continuum object.

\section{Discussion}

\subsection{Absorption Clouds}

We have detected the formaldehyde distributions in NGC~7538 and G29.96 though observations of the 2~cm and 6~cm lines in absorption against the \ion{H}{2} regions' continuum radiation.
The absorbing systems offer information about the ambient formaldehyde environment.
Appendix A describes the calculations which permit comparisons between the 6~cm and 2~cm lines relating to Table~9.
Section 4.2 discusses the formaldehyde calculations permitted by the masers lines.

The detection of both 6~cm and 2~cm absorption lines toward the G29.96 hot core at $v_{\rm LSR} \simeq 107$~\kms\ (Fig.~12a,b) permits a calculation of the excitation temperatures as described in Appendix~A.
The ratio of optical depths is 2.35 for which $T_{21}/T_{43} = 0.21$.
If we choose excitation temperatures in this ratio for the $N/T_x$ values listed for the hot core in Table~9, we find comparable column densities for the 6~cm and 2~cm lines, indicating that the absorbing cloud is smaller than the 6~cm beam.

The NGC~7538 masers' two velocity components show 2~cm absorption at both velocities (Fig.~11b,d).
Though the masers only exist in a region of diameter $\sim 30$~AU (Fig.~4), the clouds responsible for the 2~cm absorption lines have larger sizes comparable to the beam ($\sim 3000$~AU).
The two maser velocity components appear to come from two separate cloud environments.
In addition to (1) the detection of the of the distinct absorbing velocities, evidence for two separate NGC~7538 masers includes:
(2) the components have disparate brightness temperatures ($10^8$~K compared to $10^6$~K) and maser gains ($-11$ compared to $-6$),
(3) the components are not connected by the velocity gradient observed in the maser gas (Fig.~5),
and (4) the masers appear to increase intensity independently (Fig.~8).
We conclude that the masers occur in separate clouds.

\subsection{Maser Gain Calculation}

The maser amplification process depends on the column density of the maser species.
In this section we calculate the formaldehyde column densities from the observed maser gains.
We use the level notation from Appendix~A for our calculation.
We find that a comparison between the column densities derived from absorption lines and maser lines indicates a relatively strong maser inversion.

The gain of the maser amplification of background radiation can be expressed as ({\it e.g}.\ Elitzur 1992)
$$ \tau = \frac{h \nu_{21}}{4 \pi \Delta{v}}\,g_2\,B_{21}\,\int (n_2 - n_1) dl\ , $$
where $B_{21}$ is the Einstein $B$-coefficient for stimulated emission and $l$ is the amplification path or gain length.
Here we express $\tau$ in the form of Equation~21 from Boland \& de Jong (1981),
$$ \tau = 3.4 \times 10^{-13}\,\eta\,x({\rm H}_2{\rm CO})\, n({\rm H}_2)\,\frac{\Delta{l}}{\Delta{v}}\ , $$
where $\eta = -(n_2 - n_1)/n({\rm H}_2{\rm CO})$ such that a more negative number is a stronger inversion, $x({\rm H}_2{\rm CO}) \cdot n({\rm H}_2) = n({\rm H}_2{\rm CO})$, $x({\rm H}_2{\rm CO})$ is the fractional abundance of ortho-formaldehyde with respect to the total density of hydrogen, and $n({\rm H}_2)$ is the density of hydrogen.
The values of the parameters used in our calculations are listed in Table~10.

For the values obtained from VLBA observations ($\tau \simeq -12$ and $\Delta{v} = 0.5$~\kms, see Table~5), we find $\eta \cdot x({\rm H}_2{\rm CO}) \cdot n({\rm H}_2) \cdot \Delta{l} = -1.8 \times 10^{13}\ {\rm cm}^{-2}$.
We derive limiting values of $\eta$ from the following arguments:
(1) We consider $\Delta{l}$ to be geometrically constrained to be comparable to the radius of the \ion{H}{2} region, $R_{II}$.
For $\Delta{l} = R_{II} = 2.8 \times 10^{16}\ {\rm cm}$, $\eta \cdot x({\rm H}_2{\rm CO}) \cdot n({\rm H}_2) = -6.3 \times 10^{-4}\ {\rm cm}^{-3}$.
(2) For NGC~7538-IRS1, van der Tak et al.\ (2000) have determined $x({\rm H}_2{\rm CO}) = 10^{-8}$.
Adopting this value, we obtain $\eta \cdot n({\rm H}_2) = -6.3 \times 10^{4}\ {\rm cm}^{-3}$.
(3) Any non-thermal H$_2$CO level populations must occur in the density range $n({\rm H}_2) < 6 \times 10^5\ {\rm cm}^{-3}$ (Green et al.\ 1978) or else the level populations `thermalize' and any pumping is quenched.
For instance, the Orion-KL emission lines originate in gas of density $n({\rm H}_2) \sim 10^7\ {\rm cm}^{-3}$ which could not support non-thermal excitation (Kunter \& Thaddues 1971; Johnston et al.\ 1983).
For $n({\rm H}_2) < 6 \times 10^5\ {\rm cm}^{-3}$ we obtain $\eta < -0.1$.
The maser population inversion is defined not to exceed unity such that $\eta > -1$ which implies $n({\rm H}_2) > 6.3 \times 10^{4}\ {\rm cm}^{-3}$.
The limits $-1 < \eta < -0.1$ may be compared with the typical inversion strength of the Boland \& de Jong radiative pump model $\eta = -0.008$.

Alternately, we may assume that the column density $N = n({\rm H}_2{\rm CO}) \cdot \Delta{l}$ in the maser gas is the same as that independently found from our absorption observations (\S3.3.3).
From Table~9, a typical value is $N/T_x = 3 \times 10^{-13}\ {\rm cm}^{-2}\ {\rm K}^{-1}$.
For $T_x = 2$~K we find that $N = 6 \times 10^{-13}\ {\rm cm}^{-2}$.
Using $N \cdot \eta = -1.8 \times 10^{13}\ {\rm cm}^{-2}$ from above we obtain $\eta = -0.3$ which is consistent with the $\eta$ value reasoned from the maser gain.

\subsection{Inversion Mechanism}

A continuum radiation excitation mechanism capable of producing the high gains observed in the 6~cm masers would also be expected to affect the 2~cm doublet level populations:
the relatively flat Bremsstrahlung spectrum of \ion{H}{2} region plasmas should excite the 6~cm and 2~cm doublets comparably (to within an order of magnitude, see also Wadiak et al.\ 1988).
For example, the Boland \& de Jong (1981) radiative inversion model predicts a 2~cm maser for \ion{H}{2} region spectra which turnover at wavelength $\lambda_0 \lesssim 3$~cm.
However, no 2~cm maser is present (\S 3.3.1) though the spectrum of NGC~7538-IRS1 is observed to turnover in the favorable $\lambda_0 \lesssim 3$~cm range ({\it e.g}.\ Pratap et al.\ 1992; Gaume et al.\ 1995; Bloomer et al.\ 1998; Momose et al.\ 2001; Akabane et al.\ 2001).
Furthermore, since formaldehyde occurs in typical column densities around most \ion{H}{2} regions (including the maser objects, {\it e.g}.\ van der Tak et al.\ 2000; Sch\"{o}ier et al.\ 2002), there is no reason on the grounds of either continuum morphology or formaldehyde abundance to expect that only the few maser-bearing \ion{H}{2} regions should support a formaldehyde inversion.
Therefore, we suggest that the 6~cm maser inversion is due to a rare collisional excitation, not to the readily available centimeter wavelength continuum radiation which has been discussed by other authors ({\it e.g}.\ Pratap et al.\ 1992; PMS94; Mehringer, Goss, \& Palmer 1994).
The remarkable paucity of formaldehyde masers is likely a result of selective inversion conditions such as resonant interactions with species exclusive to only a few \ion{H}{2} regions.
In an attempt to find unusual conditions which can explain the presence of collisionally-pumped formaldehyde masers, we discuss the shock morphology of \ion{H}{2} regions.

Hill \& Hollenbach (1978) have modelled the dissociation wave and shock wave evolution of expanding \ion{H}{2} regions.
Our maser environments most resemble their ``Model 3'' in which the density is $n = 10^4\ {\rm cm}^{-3}$ and the \ion{H}{2} region radius is $R = 1.5 \times 10^{17}$~cm and the ambient magnetic field $|\vec{B}| = 33.5\ \mu{\rm G}$ of the model is closest to that of NGC~7538-IRS1 (Hoare, private communication; Momose et al.\ 2001).
For NGC~7538-IRS1 the model correctly predicts the observed H$_2$ line emissivity ($\sim 10^{-12}\ {\rm erg}\ {\rm s}^{-1}\ {\rm cm}^{-2}$, Bloomer et al.\ 1998; Hoban et al.\ 1991), shock velocity ($\sim 10$~\kms, {\it e.g}.\ Deharveng et al.\ 1979), and age ($3 \times 10^{4}$~yr, {\it e.g}.\ Martin 1973) for the \ion{H}{2} region.

The relevant aspect of the Hill \& Hollenbach model is that the initially faster dissociation wave is eventually overtaken by the shock wave.
The epoch at which the shock passage occurs is $7 \times 10^4 - 2 \times 10^5$~yr, in good agreement with the age of the IRS1 \ion{H}{2} region.
During the passage there exists a transition region about half-molecular in composition.
Also, the region contains transient phenomena such as non-equilibrium populations of various species such as cool molecular gas and hot excited atoms.

Evidence that these transition-shock collision regions are sufficiently rare to explain the observed paucity of formaldehyde masers is discussed by Bertoldi \& Draine (1996).
They find that for ``a broad range of parameters of interest'' (density, shock velocity, etc.) the dissociation and shock waves are merged and there is {\em no} relative evolution of the Hill \& Hollenbach type.
However, they suggest that clumps in the ambient molecular cloud are sites where the dissociation and shock waves {\em are} separate and where time-dependent effects may become important.
Therefore, we suggest that formaldehyde masers occur in the small number of \ion{H}{2} regions which possess
(1) separately evolving dissociation and shock waves,
(2) an age equal to the epoch of dissociation and shock wave passage, and
(3) a neighboring molecular cloud with ambient clumps of sufficient number density and formaldehyde column density to cause time-dependent wave effects.

We consider it possible that the formaldehyde maser inversion is a result of collisions between formaldehyde and a rare transient species within the relatively exotic conditions in the shock/dissociation-transition region.
However, since H$_2$CO collisions with relatively rare species, such as hot H atoms or excited H$_2$ ({\it e.g}.\ Green et al.\ 1978) have not been fully explored, the collision cross-sections required to formulate a quantitative model are not yet available.

\subsection{Other Masers in NGC~7538}

NGC~7538 is the closest of the three sources known to have H$_2$CO masers ($d = 3$~kpc compared with G29.96-0.02 ($d = 6$~kpc) and Sgr~B2 ($d = 8$~kpc)).
Also, the NGC~7538-IRS1 region contains a large number of maser species.
NGC~7538 is then best suited for studies targeted at understanding the H$_2$CO relationship to other maser species in close linear and angular proximity.
For example, empirical correlations among the H$_2$CO, OH, and H$_2$O maser species have prompted searches for new H$_2$CO masers ({\it e.g}.\ Forster et al.\ 1985).
In the search for a working H$_2$CO pump, an understanding of the physical conditions of IRS1 through the observation of other masers is valuable.
Studying the nearby masers that have relatively well-understood excitation mechanisms could considerably illuminate the H$_2$CO inversion environment.

Figure~15 shows the 22~GHz continuum image of IRS1 from Gaume et al.\ (1995) overlaid with the positions of other known masers in the region.
A protostellar outflow is thought to power the maser environment.
Based on 15~GHz continuum observations, Campbell (1984) suggests an edge-on protostellar disk and bipolar outflow.
A disk/outflow scenario is also supported by 22~GHz continuum and H66$\alpha$ recombination line observations (Gaume et al.\ 1995) which show a $\sim 100$~\kms\ stellar wind and apparent disk (cf.\ Jaffe \& Mart\'{\i}n-Pintado 1999).
Furthermore, based on CH$_3$OH maser observations, Minier et al.\ (1998) suggest a protostellar disk and a blueshifted southern outflow.
Alternately, a shell model has been proposed to explain the various maser positions (Dickel et al.\ 1982).
A cartoon summary of these models, including the molecular clumps that we suggest are important for the formaldehyde maser process, are included with the observational data in Figure~15.
Table~11 summarizes the observational uncertainties and physical conditions of the masers.

The $^{15}{\rm NH}_3$ (3,3) masers are the only IRS1 masers that could arise at the same density, temperature, and position as the H$_2$CO masers (Fig.~15; Table~11).
Furthermore, the velocity gradient observed in the $^{15}{\rm NH}_3$ (3,3) masers is consistent with the H$_2$CO maser velocity gradient shown in Figure~5.
The position of the non-metastable $^{14}{\rm NH}_3$ (9,6) maser in the IRS1 region (Madden et al.\ 1986) is also consistent with the $^{15}{\rm NH}_3$ and H$_2$CO masers' positions and velocity gradient.
The $^{14}{\rm NH}_3$ (9,6) maser species, like H$_2$CO, is quite rare and poorly-understood (Elitzur 1992).
Though the current position uncertainty of the NH$_3$ (9,6) maser is too large to positively associate it with the H$_2$CO, the two may be related.
The NGC~7538 NH$_3$ (9,6) maser has only been observed at a single epoch; however, the NH$_3$ (9,6) maser in W49 is known to be variable (Wilson \& Henkel 1988; Huettemeister et al.\ 1995).
Future observations of NH$_3$ in NGC~7538 could indicate (1) an accurate absolute position for the NGC~7538 $^{14}{\rm NH}_3$ (9,6) maser near the NGC~7538 H$_2$CO masers and/or (2) $^{14}{\rm NH}_3$ (9,6) maser intensity increase similar to the NGC~7538 H$_2$CO masers (Fig.~8).
Either (1) or (2) would suggest that the H$_2$CO and NH$_3$ masers arise in the same volume.

The H$_2$O masers in the region have the most disparate physical conditions from the H$_2$CO (Table~11).
However, recent observations (Hoare, private communication) show that the IRS1 H$_2$O masers appear related to the H$_2$CO masers.
MERLIN observations show 9 H$_2$O masers within 200~mas of the H$_2$CO positions.
Furthermore, the two H$_2$O masers closest in position and velocity to H$_2$CO maser component II ($v_{\rm LSR} \simeq 60.1$~\kms) have tripled in intensity in 10 years.
It is interesting to note that the H$_2$CO masers have also tripled in intensity over 10 years.
Since it is unlikely that the H$_2$O and H$_2$CO masers exist in the same volume, perhaps the two species are related by a shock interaction with a clump at $v_{\rm LSR} \simeq 60.1$~\kms.
Indeed, the shock-driven pumping scheme for H$_2$O masers (Elitzur, Hollenbach, \& McKee 1989) includes molecular clumps and may be compatible with the H$_2$CO pumping mechanism we suggest using the Hill \& Hollenbach (1978) and Bertoldi \& Draine (1996) models.

\section{Conclusions}

We have detected and imaged the unusual H$_2$CO masers with VLBI baselines for the first time showing milliarcsecond structure (tens of AUs in linear scale) and brightness temperatures in excess of $10^8$~K.
MERLIN images of the NGC~7538 H$_2$CO masers show the separation between the two velocity components to be $79 \pm 5$~mas and the components appear to arise in separate clouds.
We also measure a $1900\ {\rm km}\,{\rm s}^{-1}\,{\rm pc}^{-1}$ velocity gradient in the NGC~7538 maser gas.
Both NGC~7538 maser components have been shown to increase intensity on 5-10 year timescales despite remaining constant in intensity for the first 10 years after discovery.
We do not detect variability in the G29.96 masers on a two year timescale.

We detect the 2~cm formaldehyde line in absorption in both NGC~7538 and G29.96.
We detect no 2~cm line emission from any of the maser regions from which we conlcude that there is no inversion of the 2~cm levels.
From the good agreement between the column densities derived from the maser lines and the absorption lines, we find that the inversion of the masers is $\eta \simeq -0.3$.
We suggest that the strong but rare formaldehyde inversion process involves a collisional pumping interaction.
We posit that the environment which generates these pump conditions results from unusual shock dynamics around the \ion{H}{2} regions.

\acknowledgments

I.M.H.\ is supported by the NRAO pre-doctoral researcher program.
We thank D.\ S.\ Shepherd for helpful comments on an earlier draft.
We thank M.\ Hoare for communicating the MERLIN observations of the NGC~7538 H$_2$O masers.

\appendix

\section{Optical Depth Calculations}

The optical depth $\tau$ for thermal absorption is proportional to
$$\tau \propto \mu^2 \left( \frac{\nu}{\Delta{v}} \right) S \left( \frac{n_L}{g_L} - \frac{n_U}{g_U} \right)$$
where $\mu$ is the permanent dipole moment of the molecule, $\nu$ is the transition frequency, $\Delta{v}$ is the thermal Doppler line width, $S$ is the line strength, and $n$ and $g$ are, respectively, the number of absorbers and the statistical weights of the upper and lower levels of the transition.
For the 6~cm and 2~cm transitions of formaldehyde $S_{6cm} \approx 3/2$, $S_{2cm} \approx 5/6$, $g=3$ for both levels in the 6~cm doublet, and $g=5$ for both levels in the 2~cm doublet (see Townes \& Schawlow, 1955).

We may analyze our data using a four level system for which we use the following notation: $1_{11} \equiv 1$, $1_{10} \equiv 2$, $2_{12} \equiv 3$, and $2_{11} \equiv 4$.
We shall present our results as constraints on three excitation temperatures; $T_{21}$ between the 6~cm doublet levels, $T_{43}$ between the 2~cm doublet levels, and $T_{31}$ between the $J=2$ and $J=1$ doublets.
We examine only those absorbing systems for which we have detected both 6~cm and 2~cm absorption.
Also, we assume that the absorbing clouds are smaller than the telescope beam while the extent of the continuum radiation is not.

When both the 6~cm and 2~cm line have been observed we may compare the optical depths of the lines
\begin{eqnarray*}
\frac{<\tau_{\rm 2cm}>}{<\tau_{\rm 6cm}>} & = & \frac{\Omega_{6cm}}{\Omega_{2cm}} \, \frac{5}{6} \, \frac{2}{3} \, \frac{3}{5} \, \frac{N_3 \left( 1-e^{-h \nu_{43} / k T_{43}} \right)}{N_1 \left( 1-e^{-h \nu_{21} / k T_{21}} \right)} \\
                                          & \approx & \frac{\Omega_{6cm}}{\Omega_{2cm}} \, \frac{1}{3} \, \frac{\nu_{43}}{\nu_{21}} \, \frac{N_3}{N_1} \, \frac{T_{21}}{T_{43}}\ ,
\end{eqnarray*}
where it is assumed that $h \nu \ll k T$, where $\Omega$ is the solid angle of the telescope beam over which the absorption measurement is averaged, and where
$$ \frac{N_3}{N_1} = \frac{g_3}{g_1} e^{-h \nu_{31} / k T_{31}} $$
can be substituted since the total populations between the 6~cm and 2~cm doublets are approximately thermally distributed.
Substituting $\nu_{43} / \nu_{21} \simeq 3$, we have the relation
$$ \frac{<\tau_{\rm 2cm}>}{<\tau_{\rm 6cm}>} \simeq \frac{\Omega_{6cm}}{\Omega_{2cm}} \, \frac{5}{3} \, \frac{T_{21}}{T_{43}} \, e^{-h \nu_{31} / k T_{31}}\ .$$
For calculations we will use $\Omega_{6cm}/\Omega_{2cm} = 9.35$, $T_{31} = 20$~K, and $\nu_{31} = 140.84$~GHz.

\clearpage

\begin{figure}
\plotone{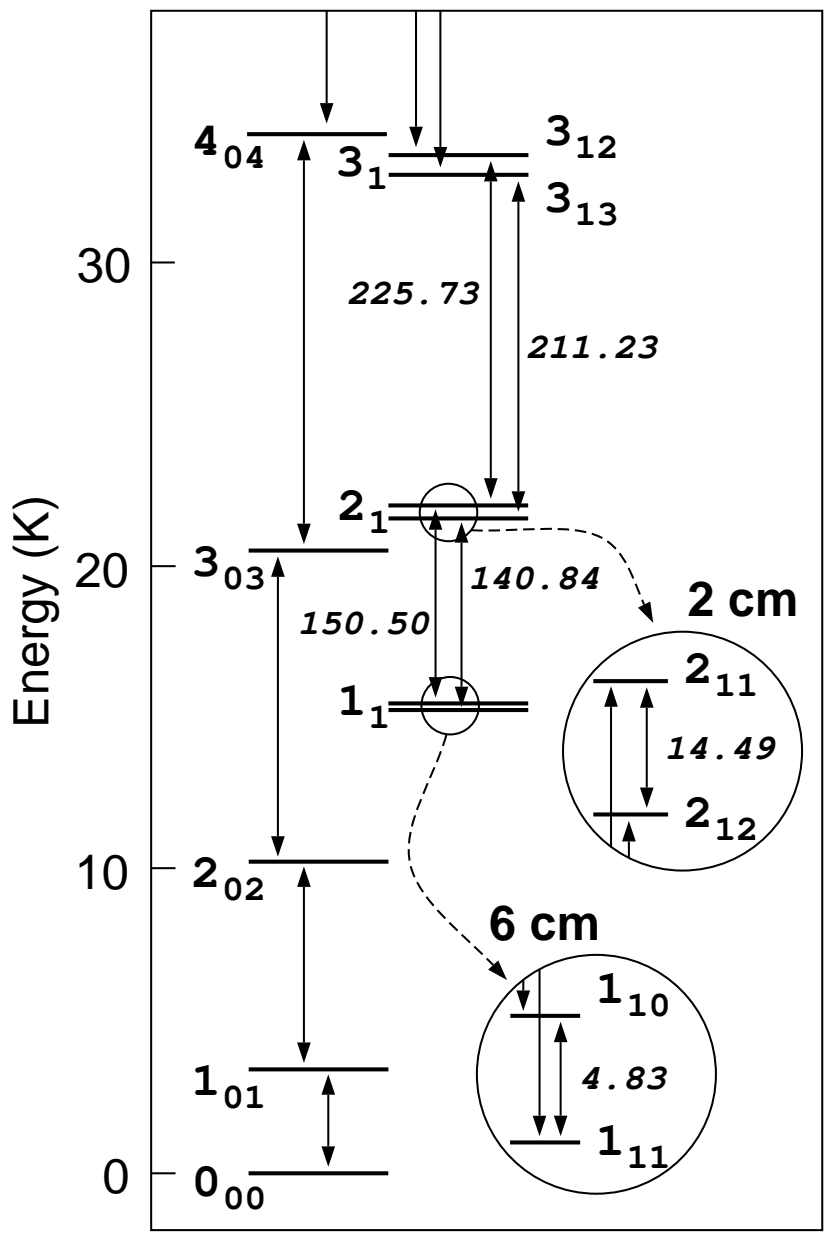}
\caption{Energy level diagram of the lowest rotational states in the ground vibrational state of formaldehyde (H$_2$CO).
The level notation is $J_{K^{-}K^{+}}$.
The $K^{-} = 1$ level splitting is exaggerated for presentation.
The solid arrows show the radiative electric dipole ($\Delta{K^{-}} = 0$, $\Delta{K^{+}} = \pm1$) transitions between the states.
Note that $K^{-}$-even and $K^{-}$-odd ladders are not radiatively connected; the $K^{-} = 0$ ladder is shown for perspective but does not enter our calculations.
The $K^{-} = 2$ and $K^{-} = 3$ ladders which begin at $E/k \simeq 60$~K and $E/k \simeq 120$~K, respectively, are not important for this paper.
The numbers in italics label the transition energies in the $K^{-} = 1$ ladder in frequency units of gigaHertz.
The insets magnify the 6- and 2-centimeter transitions of interest.
\label{fig1}}
\end{figure}

\clearpage

\begin{figure}
\plotone{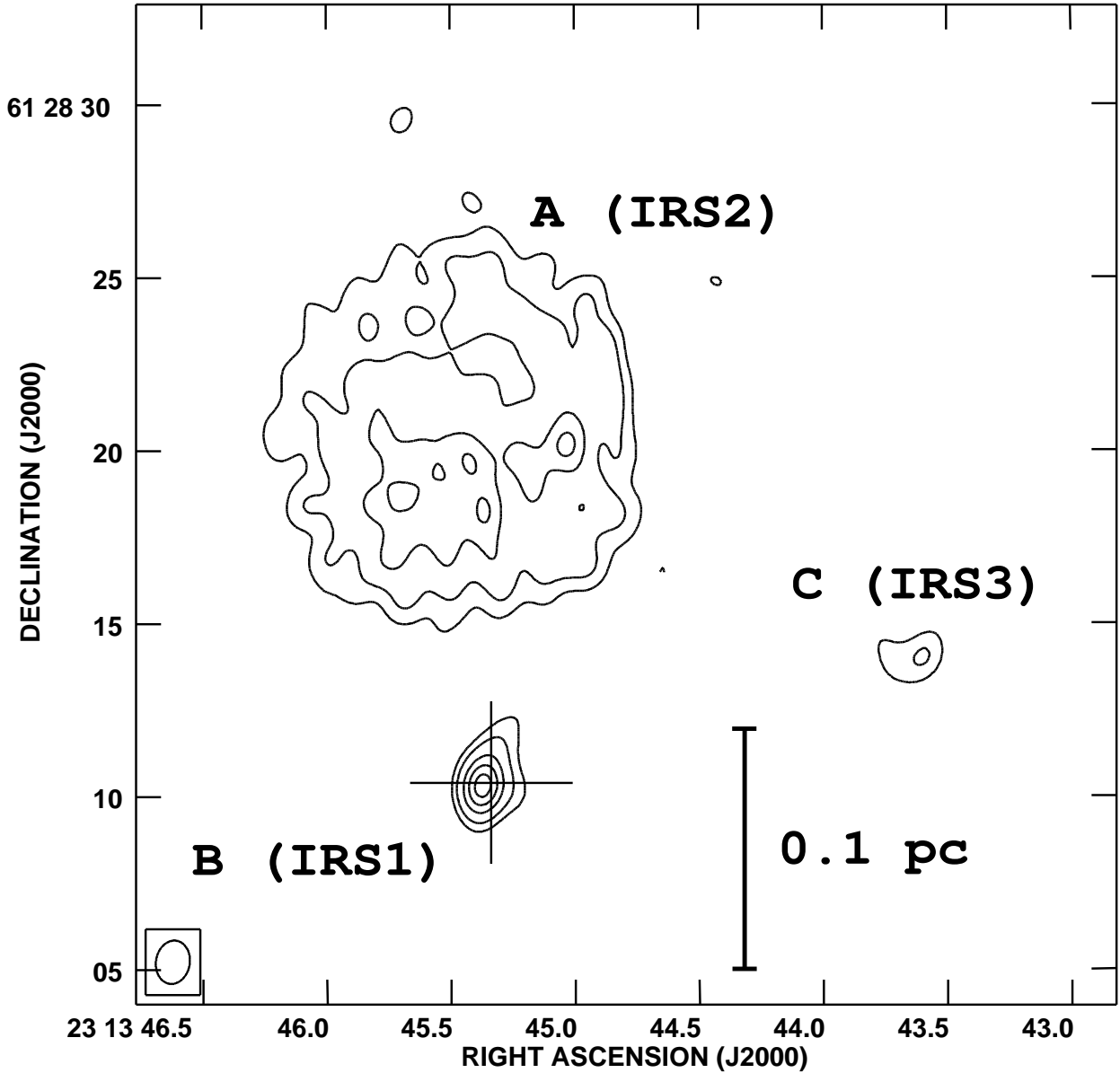}
\caption{The 4.8~GHz continuum radiation observed from NGC~7538 with the VLA `B' array.
The contours are -4, 4, 8, 16, 24, and 32 times the noise RMS 1.6~\mjb.
The beam, which is plotted in the lower left hand corner, is 1.3$\times$1.0~arcseconds at a position angle of $-10\arcdeg$.
The crossed lines mark the position of the H$_2$CO masers.
\label{fig2}}
\end{figure}

\clearpage

\begin{figure}
\plotone{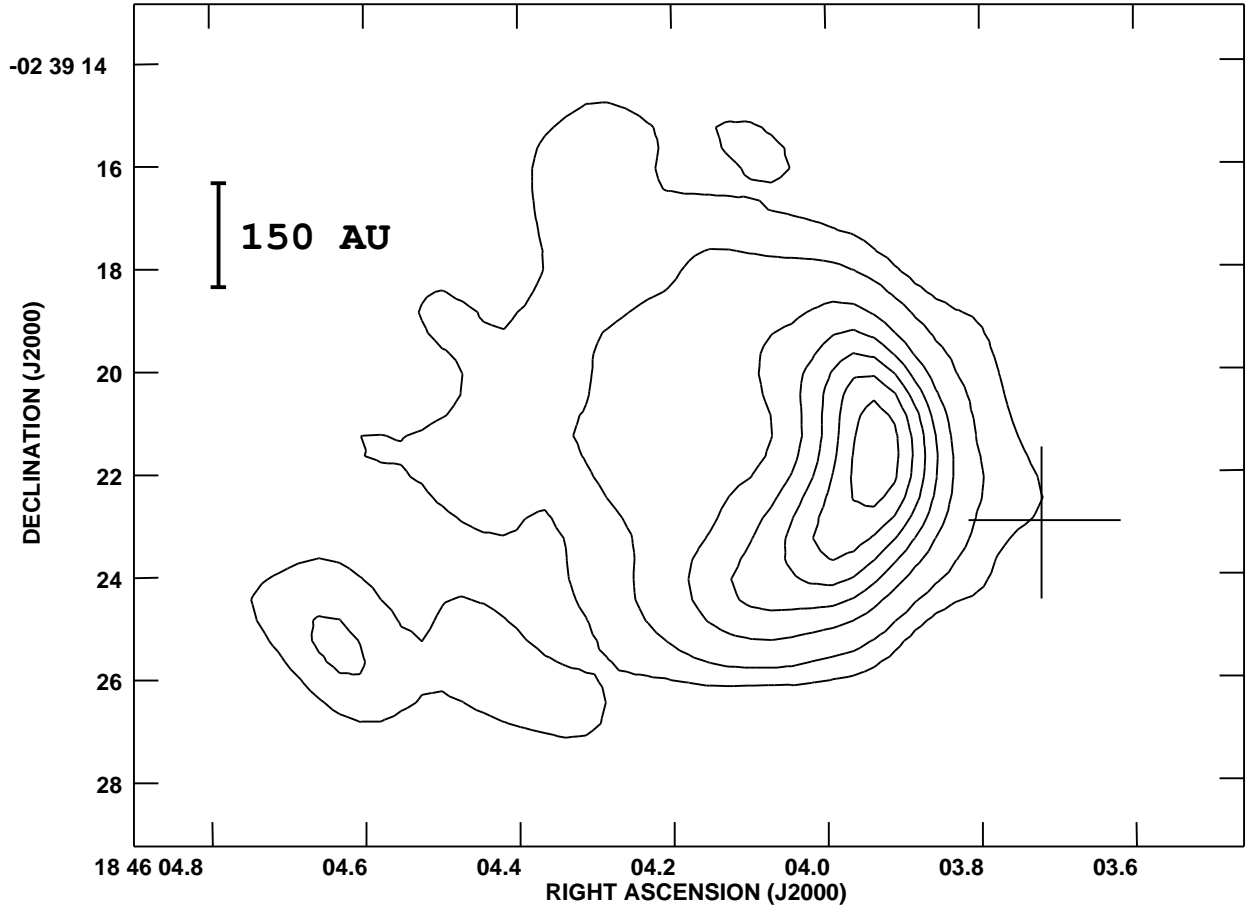}
\caption{The 14.5~GHz continuum radiation observed from G29.96-0.02 with the VLA `CnB' array.
The contours are -5, 5, 10, 20, 30, 40, 50, and 60 times the noise RMS 3.0~\mjb\ (no contours at -5 appear).
The beam is 2.0$\times$1.4~arcseconds at a position angle of $37\arcdeg$.
The crossed lines mark the position of the H$_2$CO masers.
\label{fig3}}
\end{figure}

\clearpage

\begin{figure}
\plotone{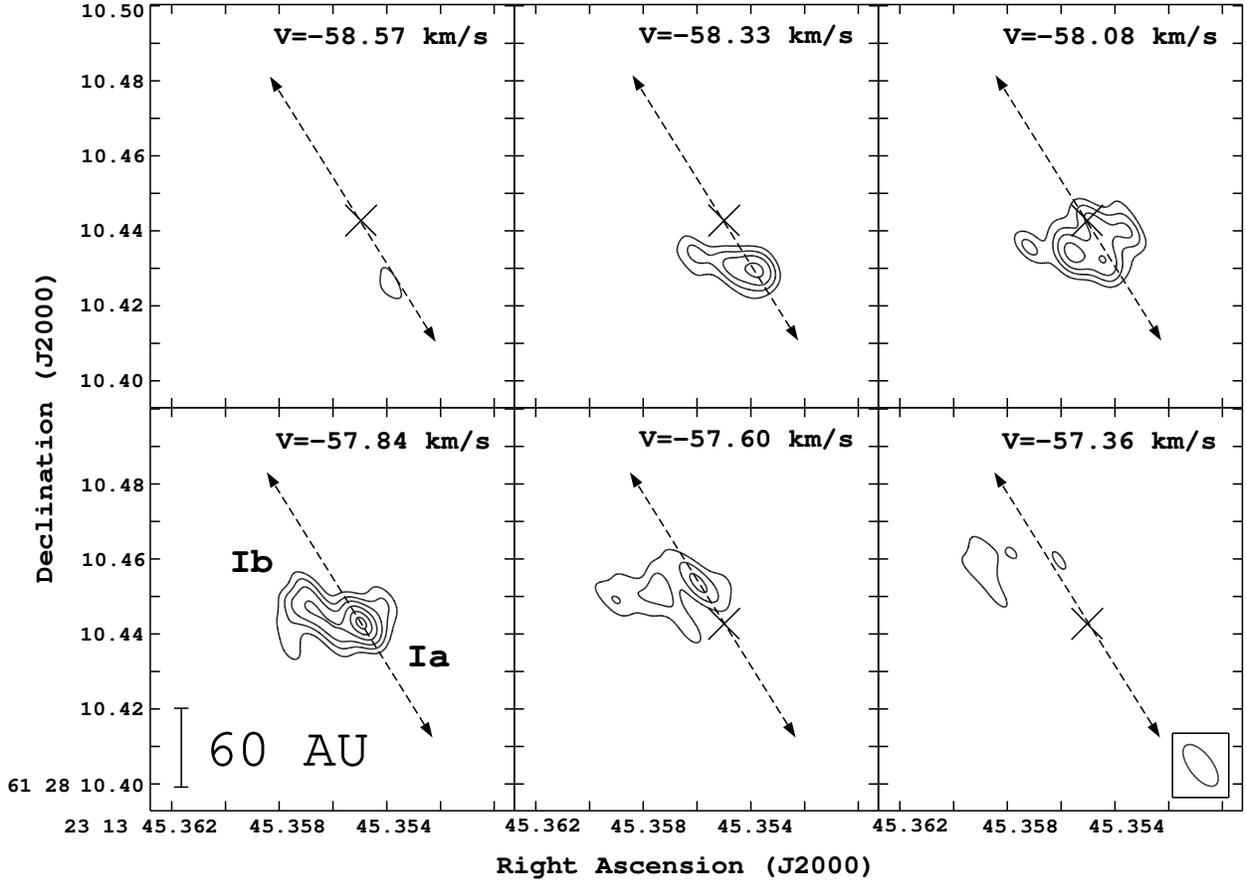}
\caption{Six adjacent channel images from the VLBA+Y27 observations of NGC~7538 maser component I.
The contours are -5, 5, 7, 9, 11.5, 14, and 16 times the noise RMS 12.8~\mjb\ (no contours at -5 appear).
The beam, which is plotted in the lower right hand corner, is 13$\times$6~mas at a position angle of $37\arcdeg$.
The dashed arrow is at position angle $31\arcdeg$, marking the direction of the linear velocity field fitted to Figure 6.
The `X' marks the position corresponding to the bright Ia peak in the $v_{\rm LSR} = -57.84$~\kms\ channel image.
Properties of the emission features are summarized in Table~4.
\label{fig4}}
\end{figure}

\clearpage

\begin{figure}
\plotone{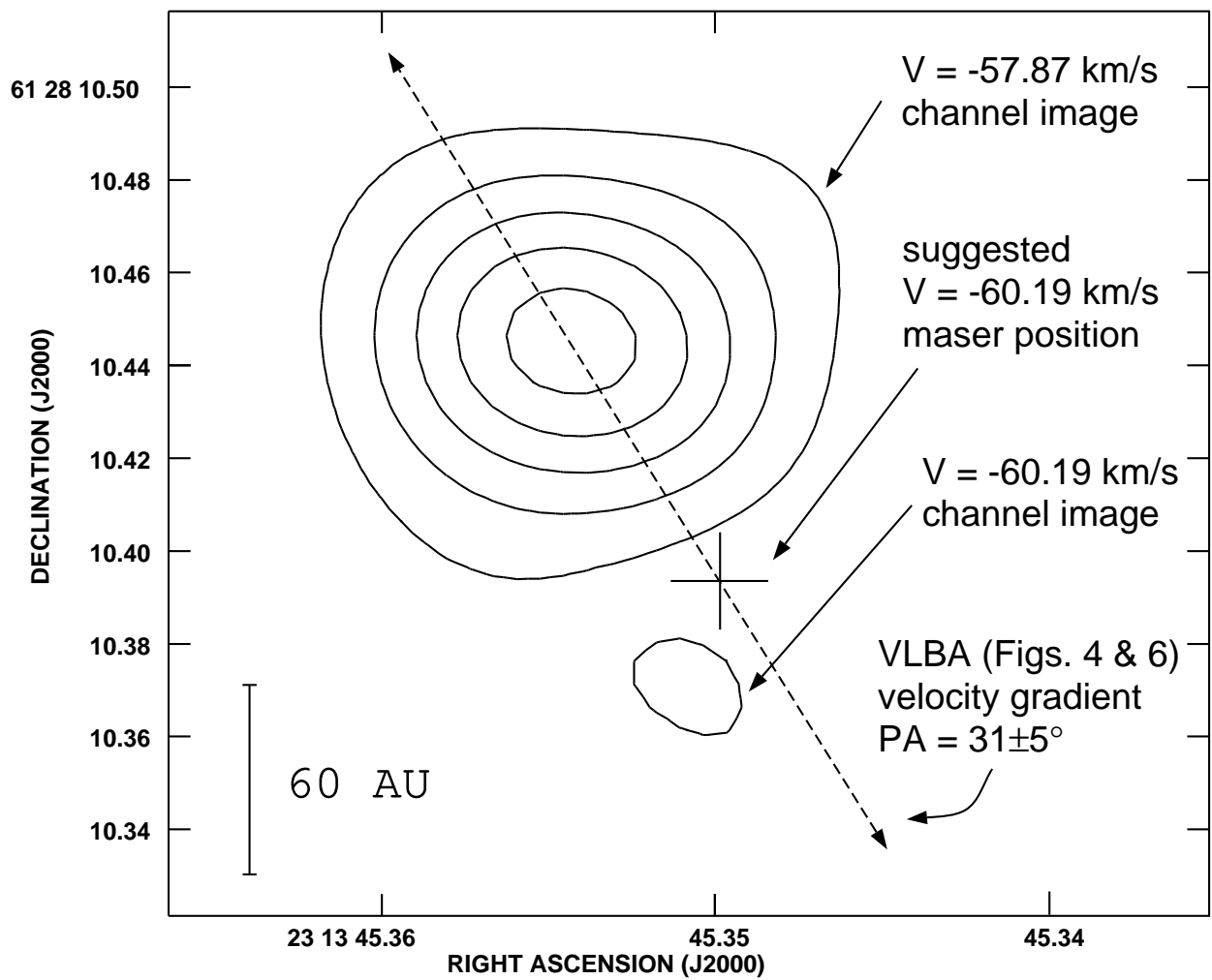}
\caption{MERLIN images of the component I and component II H$_2$CO maser line emission from NGC~7538.
The contour levels are 25, 50, 80, 100, and 130 times the image RMS noise 9~\mjb.
The beam for the MERLIN images is $53\times50$~mas at position angle $39{\arcdeg}$.
The dashed arrow represents the direction of the velocity gradient apparent in the VLBA data (Figure~4).
The `$+$' symbol marks the position at which a $v_{\rm LSR} = -60.19$~\kms\ maser would lie if consistent with the extrapolated VLBA velocity gradient (Figs.~4 \& 6).
\label{fig5}}
\end{figure}

\clearpage

\begin{figure}
\plotone{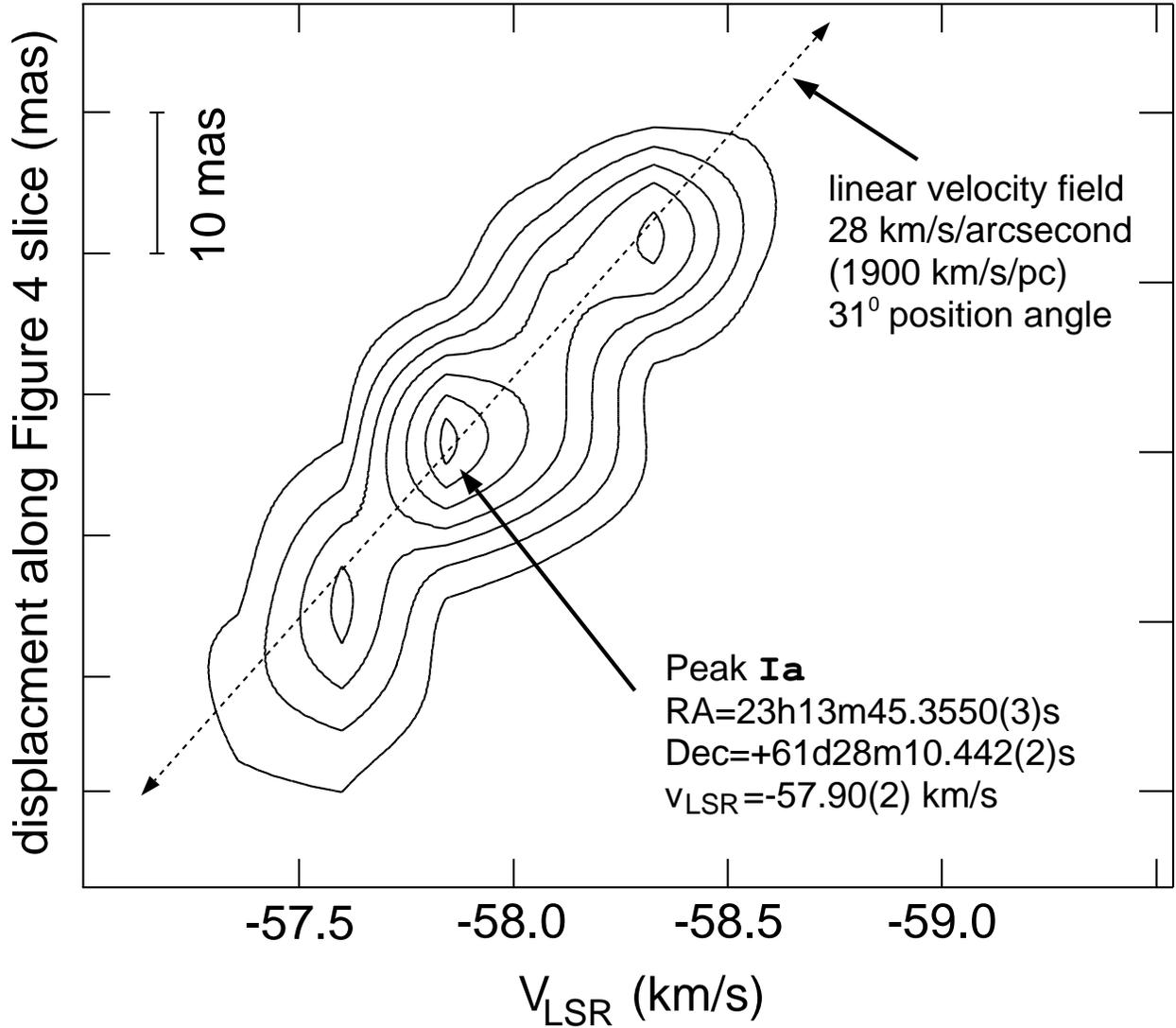}
\caption{Position-velocity plot from the VLBA+Y27 observations of NGC~7538 feature Ia showing the plane marked by the dashed arrow in Figure 4.
The contour levels are the same as in Figure 4.
The emission is fit by a linear velocity field of $1900\ {\rm km}\,{\rm s}^{-1}\,{\rm pc}^{-1}$ as shown by the dotted arrow.
\label{fig6}}
\end{figure}

\clearpage

\begin{figure}
\plotone{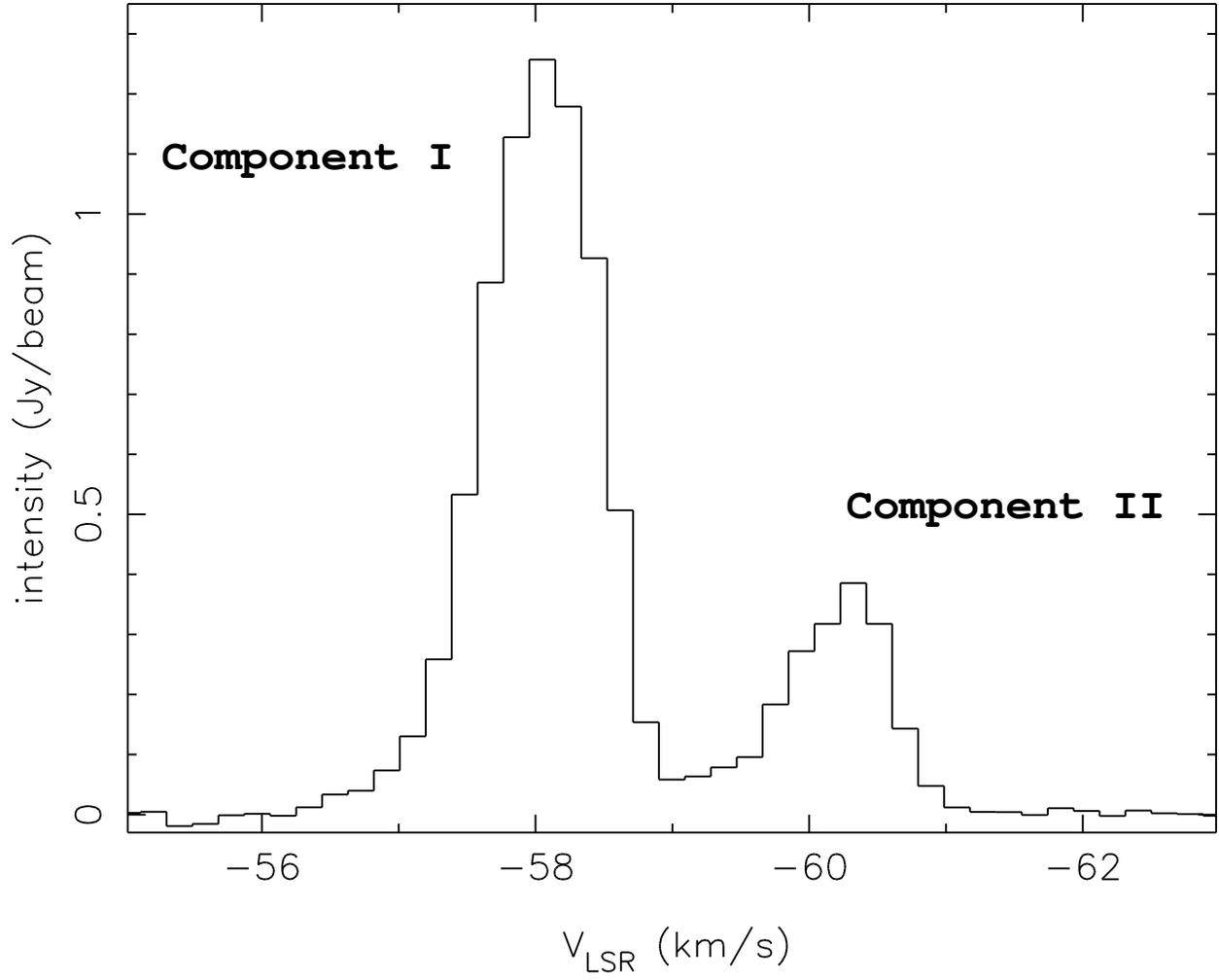}
\caption{VLA `B' array Stokes $I$ spectrum of the NGC~7538 H$_2$CO masers from July 2002 after continuum subtraction.
\label{fig7}}
\end{figure}

\clearpage

\begin{figure}
\plotone{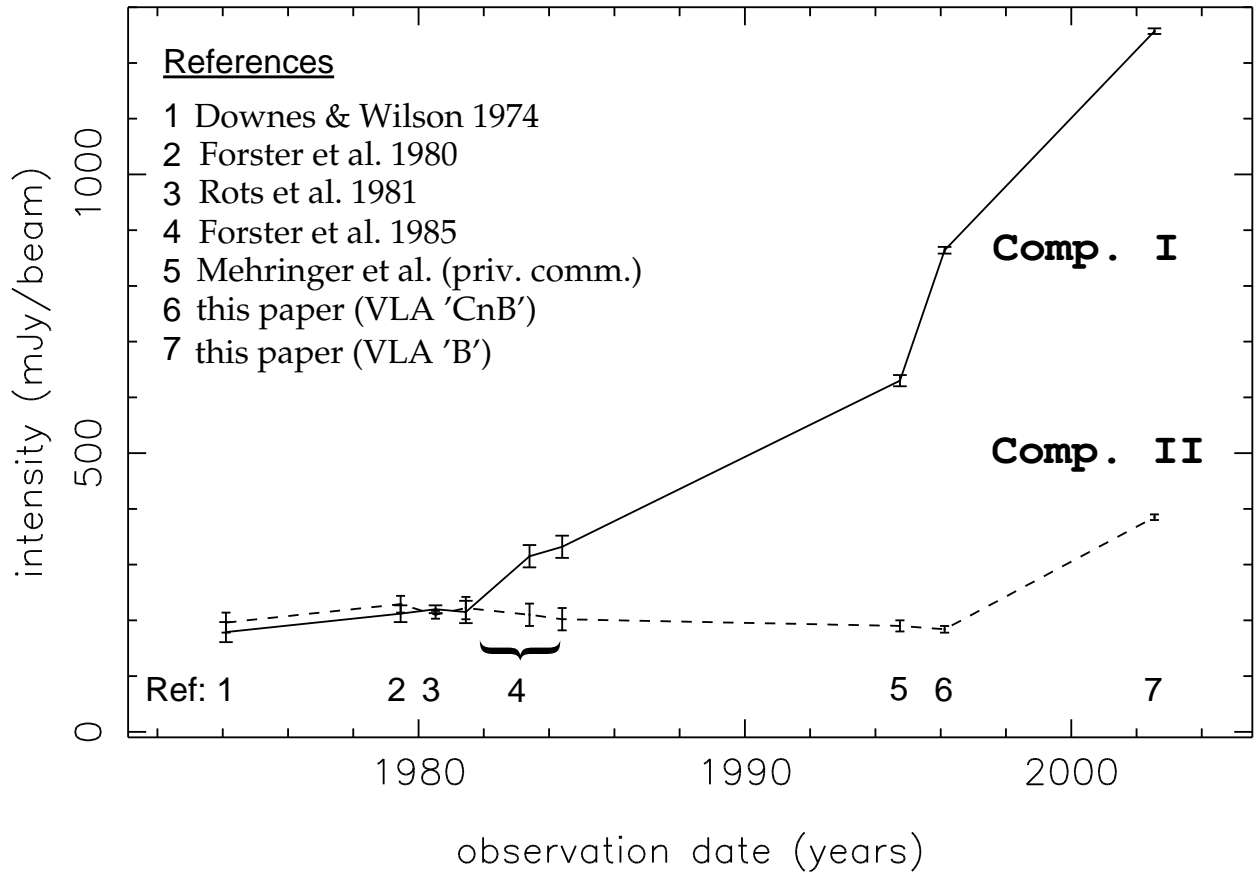}
\caption{Intensity history of the NGC~7538 H$_2$CO masers.
From the current paper, only the July 2002 VLA `B' and February 1996 VLA `CnB' observations are plotted.
\label{fig8}}
\end{figure}

\clearpage

\begin{figure}
\plotone{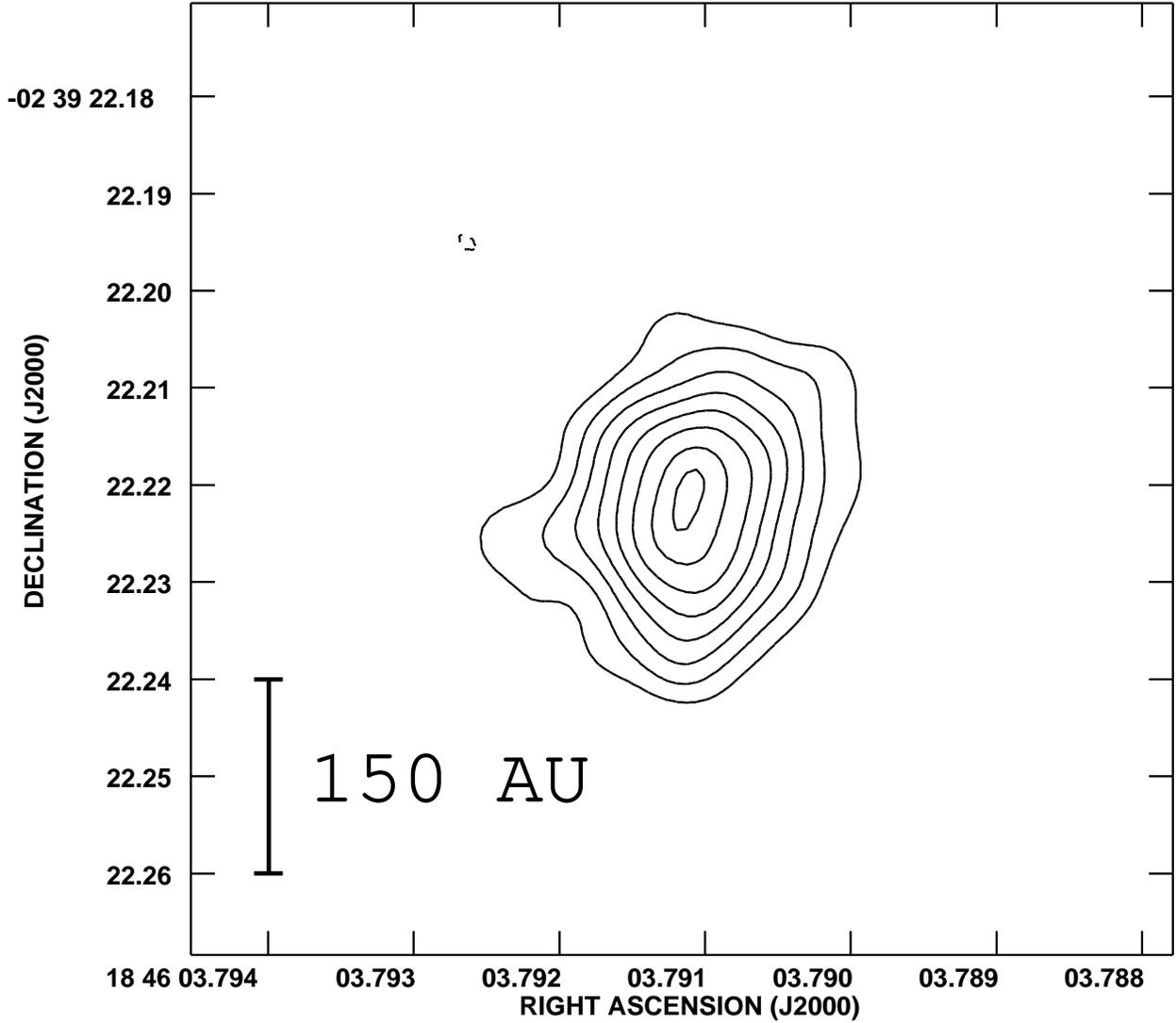}
\caption{VLBA+Y27 image of the $v_{\rm LSR} = 100.2$~\kms\ channel (component I) from the G29.96-0.02 observations.
The contours are -3, 2, 3, 4, 5, 6, 7, 8, and 8.6 times the noise RMS 7~\mjb.
The image of component II ($v_{\rm LSR} \simeq 102$~\kms) appears at the same sky position.
The beam is $23 \times 17$~mas at a position angle of $-17\arcdeg$.
The images properties of the masers are summarized in Table~6.
\label{fig9}}
\end{figure}

\clearpage

\begin{figure}
\plotone{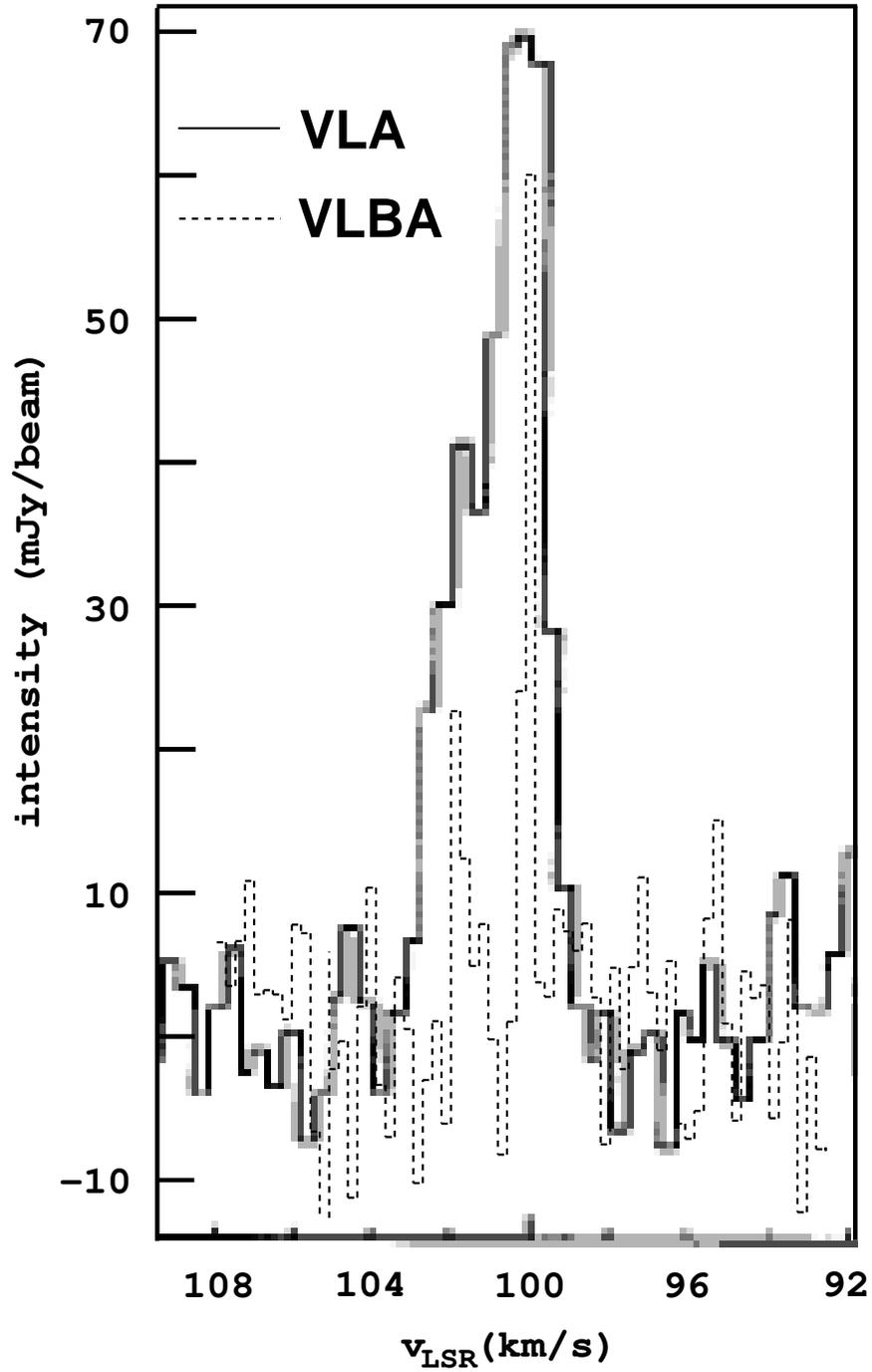}
\caption{Spectra of the H$_2$CO maser in G29.96-0.02 as observed with the VLA `A' array by Pratap et al.\ (PMS94) (solid line) and the VLBA+Y27 (dashed line).
For the Pratap et al.\ VLA observations the spectral resolution is 0.38~\kms\ and the beam size is 550~mas (0.02~pc).
For the VLBA observations the spectral resolution is 0.30~\kms\ and the beam size is 20~mas (0.0006~pc).
\label{fig10}}
\end{figure}

\clearpage

\begin{figure}
\plotone{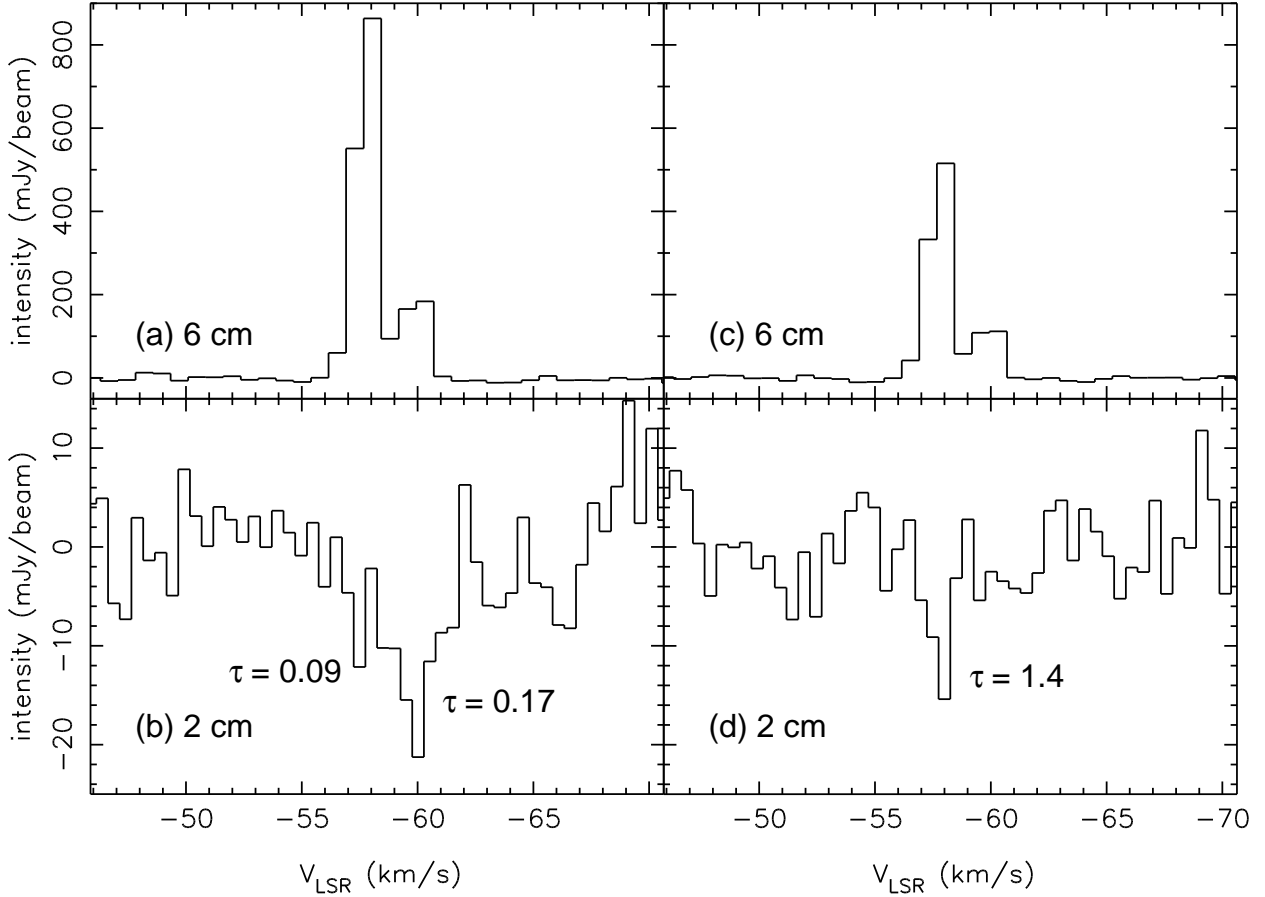}
\caption{VLA `CnB' spectra at 6~cm and 2~cm of the NGC~7538 H$_2$CO masers after continuum subtraction.
Spectra {\it (a)} and {\it (b)} are both taken at the position of the maser maximum which is also the position of deepest 2~cm absorption ($\alpha_{\rm J2000} = 23\ 13\ 45.37$, $\delta_{\rm J2000} = 61\ 28\ 10.4$).
Spectra {\it (c)} and {\it (d)} are both taken at the position of the $v_{\rm LSR} \simeq -58$~\kms\ (component I) 2~cm minimum ($\alpha_{\rm J2000} = 23\ 13\ 45.71$, $\delta_{\rm J2000} = 61\ 28\ 10.7$).
The beam size for the 6~cm observations is 5$\arcsec$ (0.07~pc), for the 2~cm observations is 2$\arcsec$ (0.03~pc).
The $\tau$ values are the peak optical depths of the respective lines.
\label{fig11}}
\end{figure}

\clearpage

\begin{figure}
\plotone{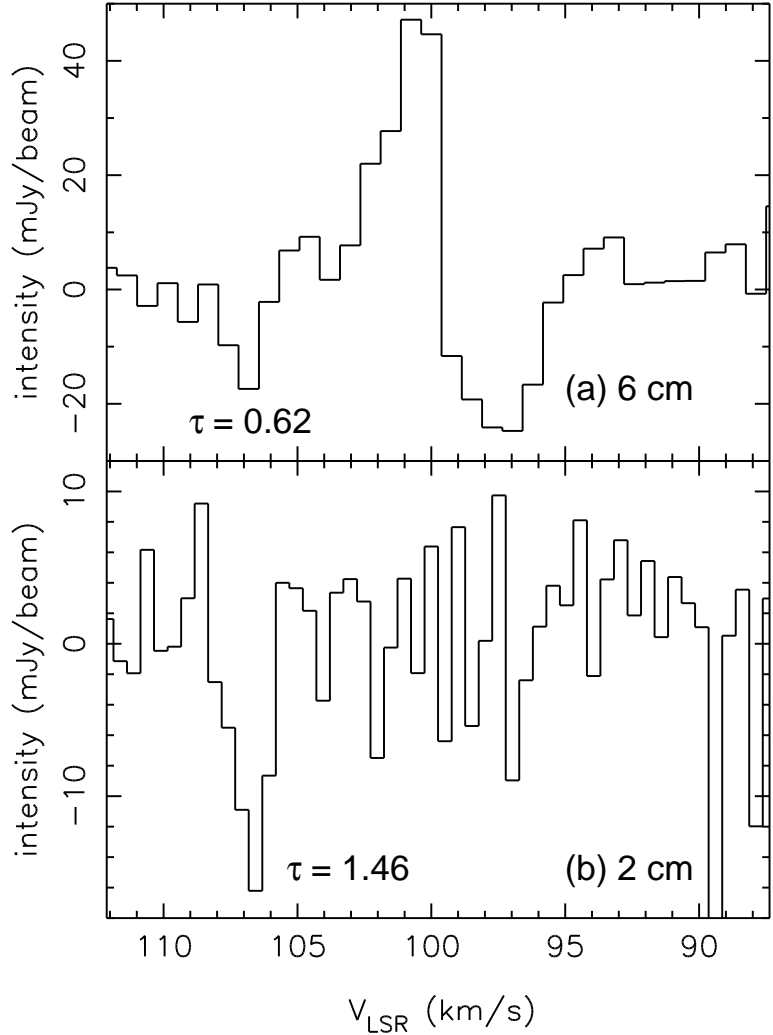}
\caption{VLA `CnB' spectra at 6~cm and 2~cm of the G29.96-0.02 H$_2$CO masers after continuum subtraction.
Spectra {\it (a)} and {\it (b)} are taken at the same sky position ($\alpha_{\rm J2000} = 18\ 46\ 03.73$, $\delta_{\rm J2000} = -02\ 39\ 22.5$): the location of the 6~cm maser maximum.
The absorption in {\it (a)} at $v_{\rm LSR} \simeq 98$~\kms\ and $v_{\rm LSR} \simeq 107$~\kms\ is subtracted in Figure~13b.
The beam size for the 6~cm observations is 5$\arcsec$ (0.16~pc), for the 2~cm observations is 2$\arcsec$ (0.06~pc).
The $\tau$ values are the peak optical depths of the respective lines.
\label{fig12}}
\end{figure}

\clearpage

\begin{figure}
\plotone{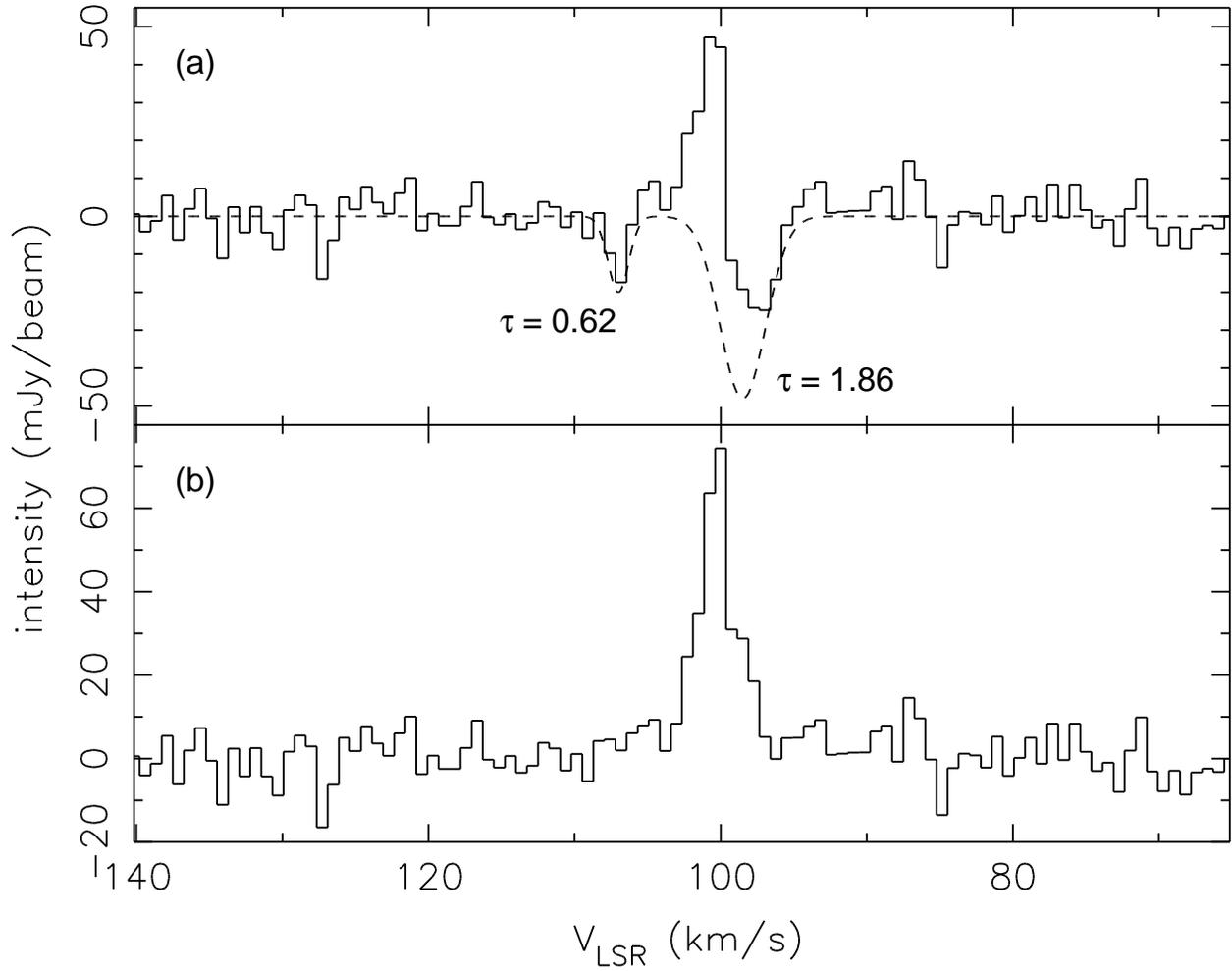}
\caption{VLA `CnB' 6~cm spectra for G29.96-0.02.
{\it (a)} shows the observed spectrum of Figure~12a (solid histogram) and the fitted absorption profiles (dotted curve).
{\it (b)} shows the spectrum in {\it (a)} after subtraction of the fitted absorption summarized in Table~8.
The $\tau$ values are the peak optical depths of the respective lines.
\label{fig13}}
\end{figure}

\clearpage

\begin{figure}
\plotone{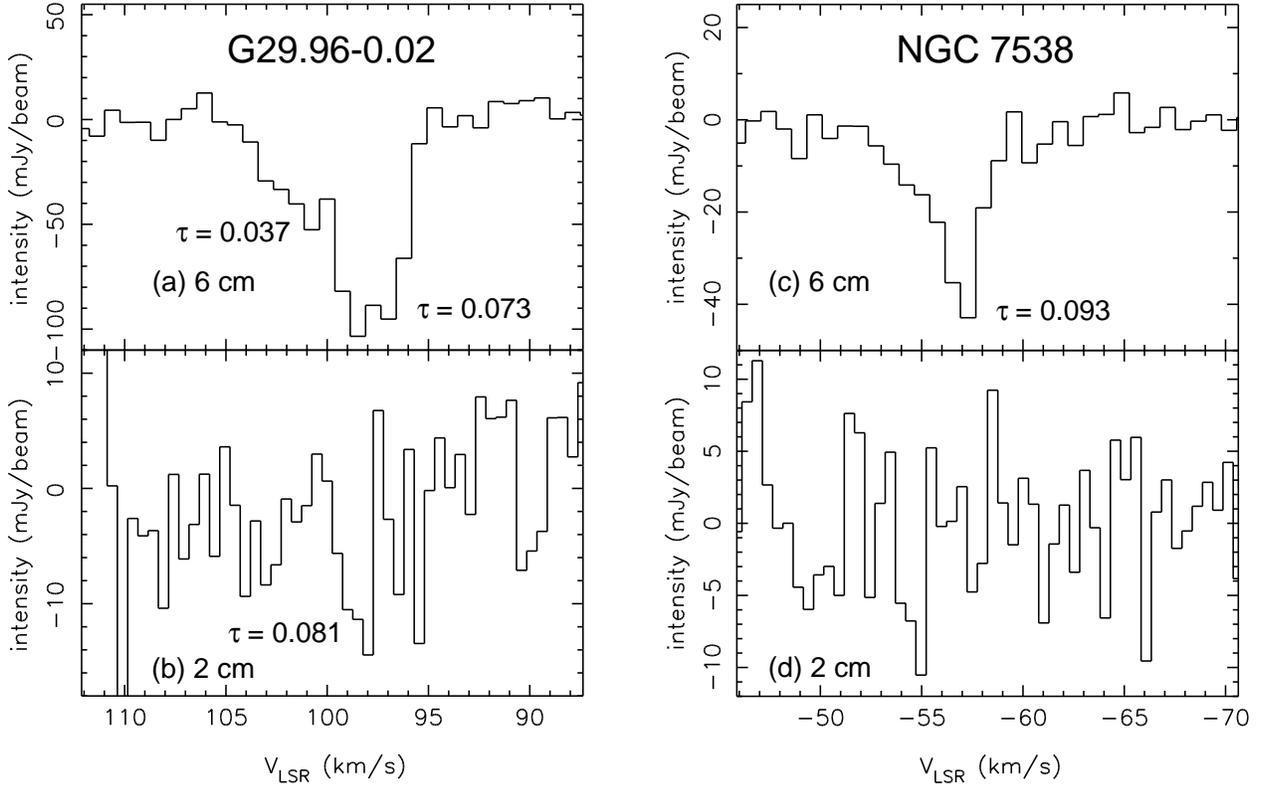}
\caption{Comparison of VLA `CnB' 6~cm and 2~cm absorption spectra.
The position of {\it (a)} \& {\it (b)} is the location of the 6~cm minimum in the G29.96-0.02 observations (6~cm beam is 5$\arcsec$ (0.16~pc), 2~cm beam is 2$\arcsec$ (0.06~pc)).
The position of {\it (c)} \& {\it (d)} is the location of the 6~cm minimum in the NGC~7538 observations (6~cm beam is 5$\arcsec$ (0.07~pc), 2~cm beam is 2$\arcsec$ (0.03~pc)).
The $\tau$ values are the peak optical depths of the respective lines.
\label{fig14}}
\end{figure}

\clearpage

\begin{figure}
\plotone{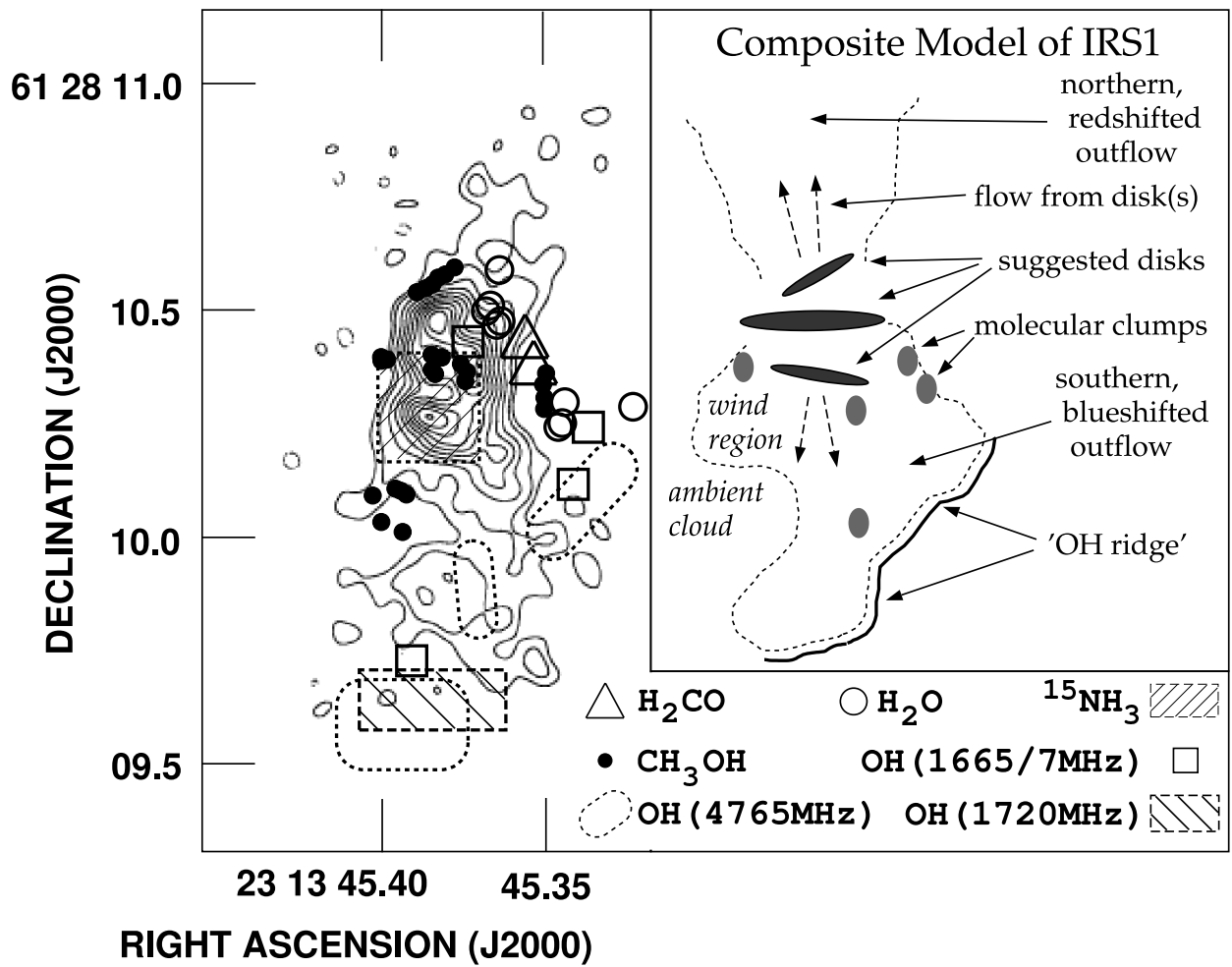}
\rotate
\caption{({\it left}) The 60-mas-resolution 22~GHz continuum VLA image of NGC~7538~B (IRS1) from Gaume et al.\ (1995) overlaid with the positions of known masers.
The contours are -4.75, 4.75, 9.5, 14.25, 19, 23.75, 28.5, 33.25, 38, and 42.25 times the image RMS noise 0.4~\mjb.
The $^{15}{\rm NH}_3$, OH (4765~MHz), and OH (1720~MHz) masers are not plotted individually, but are shown as regions.
The absolute position uncertainty of the continuum image is 100~mas.
Due to large positional uncertainties, the  $^{14}{\rm NH}_3$ and OH (6035~MHz) masers which may lie in this region are not shown.
Positional uncertainties, physical conditions, and references for the masers are in Table~11.
({\it right}) Cartoon summary of the proposed models for IRS1.
\label{fig15}}
\end{figure}

\clearpage

\begin{deluxetable}{ c r@{.}l }
\tablecolumns{3}
\tablecaption{Observed Rest Frequencies \label{tbl-1}}
\tablewidth{0pt}
\tablehead{
\colhead{Transition} & \multicolumn{2}{c}{Frequency}
}
\startdata
$1_{10} \rightarrow 1_{11}$ (6~cm) & 4&829657~GHz \\
$2_{11} \rightarrow 2_{12}$ (2~cm) & 14&488475~GHz \\
\enddata
\end{deluxetable}

\clearpage

\begin{deluxetable}{ l c c c c c c c c c c }
\rotate
\tabletypesize{\scriptsize}
\tablecolumns{11}
\tablecaption{Observational Parameters \label{tbl-2}} 
\tablewidth{0pt}
\tablehead{
\colhead{Array} & \colhead{Band} & \colhead{$v_{\rm LSR}$} & \colhead{bandwidth} & \colhead{bandwidth} & \colhead{$n_{\rm chan}$} & \multicolumn{2}{c}{amp/fringe/band cal'r\tablenotemark{a}} & \multicolumn{3}{c}{phase calibrator} \\
                &                & \colhead{(\kms)}        & \colhead{(\kms)}      & \colhead{(kHz)}         &                          & \colhead{Source} & \colhead{$S$ (Jy)} & \colhead{Source} & \colhead{$S$ (Jy)} & \colhead{$\sigma_{\theta}$ (mas)\tablenotemark{b}}
}
\startdata
\cutinhead{NGC~7538}
VLA `CnB'   & 2cm &  -58.0 & 32.0 & 1562.5 &  63 & 3C286     & 3.40 & B2229+695   & 1.1  & $<$2 \\
VLA `CnB'   & 6cm &  -58.0 & 95.0 & 1562.5 & 127 & 3C286     & 7.47 & B2229+695   & 1.1  & $<$2 \\
VLA `B'     & 6cm &  -59.0 &  7.0 &   98.0 &  63 & 3C48      & 5.48 & B2229+695   & 1.1  & $<$2 \\
MERLIN      & 6cm &  -59.0 & 30.0 &  500.0 & 256 & 3C286     & 7.47 & B2300+638   & 1.0  & 100  \\
VLBA+Y27    & 6cm &  -59.0 & 30.0 &  500.0 & 128 & B1749+096 & 2.2  & J2302+6405  & 0.25 & $<$2 \\
\cutinhead{G29.96-0.02}
VLA `CnB'   & 2cm &  100.0 & 32.0 & 1562.5 &  63 & 3C286     & 3.40 & B1749+096   & 2.1  & $<$2 \\
VLA `CnB'   & 6cm &  100.0 & 95.0 & 1562.5 & 127 & 3C286     & 7.47 & B1821+107   & 1.3  & $<$2 \\
VLBA+Y27    & 6cm &  100.3 & 30.0 &  500.0 & 128 & B1749+096 & 2.2  & J1851+0035  & 0.4  &  10  \\
\enddata
\tablenotetext{a}{The VLA and MERLIN use this continuum source to set the amplitude scale and to characterize the bandpass. The VLBA uses this source to search for the clock-like station delays which produce fringes and to characterize the bandpass.}
\tablenotetext{b}{The absolute position uncertainty of the phase calibration source as listed in the VLA Calibrator Manual at {\tt http://www.aoc.nrao.edu/${\mathtt \sim}$gtaylor/calib.html}.}
\end{deluxetable}

\clearpage

\begin{deluxetable}{ l c c c c c c r }
\tablecolumns{8}
\tablecaption{Image Parameters \label{tbl-3}}
\tablewidth{0pt}
\tablehead{
\colhead{Array} & \colhead{Band} & \colhead{Date} & \colhead{spec.\ res.} & \colhead{$\Delta{I}_{\rm RMS}$} & \colhead{beam} & \colhead{beam P.A.} & \colhead{$\theta_{\rm LAS}$\tablenotemark{a}} \\
                &                &                &  \colhead{(\kms)}  &       \colhead{(\mjb)}          & \colhead{(as)}& \colhead{(deg)} & \colhead{(as)} \\
}
\startdata
\cutinhead{NGC~7538-IRS1}
VLA `CnB'       & 2cm   & 13 Feb 1996   & 0.60  & 6     & 1.9$\times$1.5	& 53	& 90 \\
VLA `CnB'	& 6cm	& 13 Feb 1996	& 0.80	& 6	& 5.7$\times$4.5	& 38	& 120 \\
VLA `B'		& 6cm	& 24 Jul 2002	& 0.20	& 5	& 1.3$\times$1.0	& -11	& 36 \\
MERLIN		& 6cm	& 02 Oct 2001	& 0.14	& 9	& 0.053$\times$0.050	& 38	& 1.2 \\
VLBA+Y27	& 6cm	& 23 Sep 2000	& 0.30	& 13	& 0.013$\times$0.006	& 37	& 0.25 \\
\cutinhead{G29.96-0.02}
VLA `CnB'       & 2cm   & 13 Feb 1996   & 0.60  & 6     & 2.1$\times$1.5	& 22	& 90 \\
VLA `CnB'       & 6cm   & 13 Feb 1996   & 0.80	& 6 	& 6.4$\times$4.6	& 11	& 120 \\
VLBA+Y27	& 6cm	& 23 Sep 2000	& 0.30	& 7	& 0.023$\times$0.017	& -17	& 0.25 \\
\enddata
\tablenotetext{a}{largest angular scale to which the observations are sensitive}
\end{deluxetable}

\clearpage

\begin{deluxetable}{ c r r c c c c c }
\tablecaption{NGC~7538 $v_{\rm LSR}$ = -57.84~\kms\ VLBA Image Features \label{tbl-4}}
\tablewidth{0pt}
\tablehead{
\colhead{Feature} & \colhead{R.A.\tablenotemark{a}} & \colhead{Dec.\tablenotemark{a}} & \colhead{$I$} & \colhead{$\theta_{\rm max}$} & \colhead{$T_B$} \\
 & \colhead{(h m s)} & \colhead{($\arcdeg$ $\arcmin$ $\arcsec$)} & \colhead{(\mjb)} & \colhead{(mas)} & \colhead{($10^8$~K)} \\
}
\startdata
 Ia & 23 13 45.3550(3) & 61 28 10.442(2) &  216(13) & 9(3) & 1.6(2) \\
 Ib &       45.3568(3) &       10.447(2) &  153(13) & 11(4) & 1.4(2) \\
\enddata
\tablenotetext{a}{Coordinates in J2000 epoch}
\end{deluxetable}

\clearpage

\begin{deluxetable}{ r r r r }
\tablecolumns{4}
\tablecaption{NGC~7538 Maser Line Profiles \label{tbl-5}}
\tablewidth{0pt}
\tablehead{
\colhead{Feature} & \colhead{$I$} & \colhead{$v_{\rm LSR}$} & \colhead{$\Delta{v}$} \\
 & \colhead{(\mjb)} & \colhead {(\kms)} & \colhead {(\kms)} \\
}
\startdata
\cutinhead{VLA `B'}
  I & 1268(20) & -58.05(1) & 0.99(6) \\
 II &  372(20) & -60.22(8) & 0.88(8) \\
\cutinhead{MERLIN}
I & 1310(35) & -57.90(4) & 0.82(5) \\
II & 270(35) & -60.19(6) & 0.64(8) \\
\cutinhead{VLBA+Y27}
Ia & 216(13) & -57.90(2) & 0.37(9) \\
Ib & 153(13) & -57.82(2) & 0.41(9) \\
II & \tablenotemark{a} & & \\
\enddata
\tablenotetext{a}{feature not imaged}
\end{deluxetable}

\clearpage

\begin{deluxetable}{ c r r c c c c c }
\tablecaption{G29.96-0.02 VLBA Image Features \label{tbl-6}}
\tablewidth{0pt}
\tablehead{
\colhead{Feature} & \colhead{R.A.\tablenotemark{a}} & \colhead{Dec.\tablenotemark{a}} & \colhead{$I$} & \colhead{$\theta_{\rm max}$} & \colhead{$T_B$} \\
 & \colhead{(h m s)} & \colhead{($\arcdeg$ $\arcmin$ $\arcsec$)} & \colhead{(\mjb)} & \colhead{(mas)} & \colhead{($10^6$~K)} \\
}
\startdata
 I  & 18 46 03.7911(7) & -02 39 22.221(10) & 61(8) & 18(7) & 9.2(12) \\
 II &       03.7910(7) &        22.221(10) & 23(8) & 20(9) & 3.5(12) \\
\enddata
\tablenotetext{a}{Coordinates in J2000 epoch}
\end{deluxetable}

\clearpage

\begin{deluxetable}{ r r r r }
\tablecolumns{4}
\tablecaption{G29.96-0.02 Maser Line Profiles \label{tbl-7}}
\tablewidth{0pt}
\tablehead{
\colhead{Feature} & \colhead{$I$} & \colhead{$v_{\rm LSR}$} & \colhead{$\Delta{v}$} \\
 & \colhead{(\mjb)} & \colhead {(\kms)} & \colhead {(\kms)} \\
}
\startdata
\cutinhead{VLA `CnB'}
I (\& II) & 64(26) & 100.43(5) & 2.42(64) \\
\cutinhead{VLBA+Y27}
I  & 70(18) & 100.24(4) & 0.288(87) \\
II & 19(7)  & 102.0(2)  & \tablenotemark{a} \\
\enddata
\tablenotetext{a}{feature not spectrally resolved}
\end{deluxetable}

\clearpage

\begin{deluxetable}{ r r r }
\tablecolumns{3}
\tablecaption{G29.96-0.02 Hot Core 6~cm Absorption \label{tbl-8}}
\tablewidth{0pt}
\tablehead{
\colhead{$v_{\rm LSR}$} & \colhead{$\Delta{v}$} & \colhead{$I$} \\
\colhead{(\kms)}        & \colhead{(\kms)}                 & \colhead{(\mjb)}
}
\startdata
 98.20(120) & 3.4(15) & -58(22) \\
 107.11(90) & \tablenotemark{a} & \tablenotemark{a} \\
\enddata
\tablenotetext{a}{feature not spectrally resolved}
\end{deluxetable}

\clearpage

\begin{deluxetable}{ r c c c }
\tablecolumns{4}
\tablecaption{Absorption Column Densities \label{tbl-9}}
\tablewidth{0pt}
\tablehead{
\colhead{$v_{\rm LSR}$} & \colhead{$N/T_x$ (from 6cm)}               & \colhead{$N/T_x$ (from 2cm)}               & \colhead{Figure} \\
\colhead{(\kms)}        & \colhead{$(10^{13}\ {\rm cm}^{-2}\ {\rm K}^{-1})$} & \colhead{($10^{13}\ {\rm cm}^{-2}\ {\rm K}^{-1}$)} & \\
}
\startdata
\cutinhead{NGC~7538-IRS1}
-57.6 & \tablenotemark{a} & 0.27 & 11b \\
-58.1 & \tablenotemark{a} & 2.42 & 11d \\
-60.1 & \tablenotemark{a} & 3.36 & 11b \\
\cutinhead{NGC~7538-IRS2}
-57.3 & 3.60 & \tablenotemark{a} & 14c \\
\cutinhead{G29.96-0.02 Hot Core}
107.1 & 2.17 & 7.79 & 13a, 12b \\
 98.2 & 18.99 & \tablenotemark{a} & 13a \\
\cutinhead{G29.96-0.02 Cometary \ion{H}{2} Region}
100.8 & 0.71 & \tablenotemark{a} & 14a \\
 98.5 & 0.61 & 0.78 & 14a,b \\
\enddata
\tablenotetext{a}{no absorption detected}
\end{deluxetable}

\clearpage

\begin{deluxetable}{ l c c }
\tablecolumns{3}
\tablecaption{Maser Gain Parameters \label{tbl-10}}
\tablewidth{0pt}
\tablehead{
\colhead{Parameter} & \colhead{Symbol} & \colhead{Value}
}
\startdata
line width & $\Delta{v}$ & 0.5~\kms \\
gain length & $\Delta{l}$ & $2.8 \times 10^{16}\ {\rm cm}$ \\
H$_2$CO fraction & $x({\rm H}_2{\rm CO})$ & $10^{-8}$ \\
gas density & $n({\rm H}_2)$ & $6 \times 10^{4-5}\ {\rm cm}^{-3}$ \\
\enddata
\end{deluxetable}

\clearpage

\begin{deluxetable}{ l r r c  r c }
\tablecolumns{6}
\tablecaption{NGC~7538-IRS1 Maser Parameters \label{tbl-11}}
\tablewidth{0pt}
\tablehead{
\colhead{Maser}  & \multicolumn{3}{c}{Model Conditions} & \multicolumn{2}{c}{Fig.~15 Position} \\
\colhead{Species} & \colhead{$T_{kin}$} & \colhead{$\log_{10}{n}$} &  & \multicolumn{2}{c}{Uncertainty} \\
                 & \colhead{(K)}       & \colhead{(${\rm cm}^{-3}$)} & \colhead{Ref.} & \colhead{(mas)} & \colhead{Ref.}
}
\startdata
 H$_2$CO		& $ 20$		& $ <5$		& 1	& 2	& \S3.1.1	\\
 $^{15}$NH$_3$ (3,3)    & $> 50$        & ${4-5}$       & 2,3	& 100	& 10  \\
 OH (1720 MHz)		& $ 90$		& $ >5$		& 4	& 100	& 11 \\
 OH (4765 MHz)		& $\sim 90$	& $ >4$		& 5,12	& 500	& 12 \\
 OH (6035 MHz)		& $> 170$	& $ >7$		& 6	& 1800 	& 13 \\
 CH$_3$OH		& $> 50$	& ${7-8}$	& 7	& 30	& 14 \\
 H$_2$O			& $ 400$	& ${7-9}$	& 8	& 15 	& 15,16 \\
 $^{14}$NH$_3$ (9,6)	& $\sim 300$	& ${4-10}$	& 9,17	& 3000 	& 17 \\
\enddata
\tablecomments{References: (1) Boland \& de Jong 1981, (2) Walmsley \& Ungerechts 1983, (3) Flower, Offer, \& Schilke 1990, (4) Guibert, Elitzur, \& Nguyen-Q-Rieu 1978, (5) Elitzur 1976, (6) Pavlakis \& Kylafis 2000, (7) Sutton et al.\ 2001, (8) Elitzur, Hollenbach, \& McKee 1989, (9) Mauersberger, Henkel, \& Wilson 1987, (10) Gaume et al.\ 1991, (11) Forster et al.\ 1982, (12) Palmer, Gardner, \& Whiteoak 1984, (13) Baudry et al.\ 1997, (14) Minier, Booth, \& Conway 2000, (15) Kameya et al.\ 1990, (16) Hoare, priv.\ comm., (17) Madden et al.\ 1986}
\end{deluxetable}


\begin{thebibliography}{}
\bibitem[Akabane et al.\ (2001)]{aka01} Akabane, Kenji, Matsuo, Hiroshi, Kuno, Nario, \& Sugitani, Koji 2001, \pasj, 53, 821
\bibitem[Baan et al.\ (1986)]{baa86} Baan, W.\ A., G\"usten, R., \& Haschick, A.\ D.\ 1986, \apj, 305, 830
\bibitem[Baan et al.\ (1993)]{baa93} Baan, W.\ A., Haschick, A.\ D., \& Uglesich, R.\ 1993, \apj, 415, 140
\bibitem[Baudry et al.\ (1997)]{bau97} Baudry, A., Desmurs, J.\ F., Wilson, T.\ L., \& Cohen, R.\ J.\ 1997, \aap, 325, 255
\bibitem[Bertoldi \& Draine (1996)]{ber96} Bertoldi, F.\ \& Draine, B.\ T.\ 1996, \apj, 458, 222
\bibitem[Bieging et al.\ (1980)]{bie80} Bieging, J., Downes, D., Wilson, T.\ L., Martin, A.\ H.\ M., \& Guesten, R.\ 1980, \aaps, 42, 163
\bibitem[Bloomer et al.\ (1998)]{blo98} Bloomer, J.\ D., Watson, D.\ M., Pipher, J.\ L., Forrest, W.\ J., Ali, B., Greenhouse, M.\ A., Satyapal, S., Smith, H.\ A., Fischer, J., \& Woodward, C.\ E.\ 1998, \apj, 506, 727
\bibitem[Boland \& de Jong (1981)]{bol81} Boland, W.\ \& de Jong, T.\ 1981, \aap, 98, 149
\bibitem[Campbell (1984)]{cam84} Campbell, B.\ 1984, \apjl, 282, L27
\bibitem[Campbell \& Thompson (1984)]{ct84} Campbell, B.\ \& Thompson, R.\ I.\ 1984, \apj, 279, 650
\bibitem[Cesaroni et al.\ (1998)]{ces98} Cesaroni, R., Hofner, P., Walmsley, C.\ M., \& Churchwell, E.\ 1998, \aap, 331, 709
\bibitem[De~Buizer et al.\ (2002)]{deb02} De~Buizer, James M., Watson, Alan M., Radomski, James T., Pi\~{n}a, Robert K., \& Telesco, Charles M.\ 2002, \apjl, 564, L101
\bibitem[Deharveng et al.\ 1979]{deh79} Deharveng, L., Lortet, M.\ C., \& Testor, G.\ 1979, \aap, 71, 151
\bibitem[Dickel et al.\ (1982)]{dic82} Dickel, H.\ R., Rots, A.\ H., Goss, W.\ M., \& Forster, J.\ R.\ 1982, \mnras, 198, 265
\bibitem[Downes \& Wilson (1974)]{dow74} Downes, D.\ \& Wilson, T.\ L.\ 1974, \apjl, 191, L77
\bibitem[Downes et al.\ (1976)]{dow76} Downes, D., Wilson, T.\ L., \& Bieging, J.\ 1976, \aap, 52, 321
\bibitem[Elitzur (1976)]{eli76} Eliztur, Moshe 1976, \apj, 203, 124
\bibitem[Elitzur et al.\ (1989)]{eli89} Elitzur, Moshe, Hollenbach, David J., \& McKee, Christopher F.\ 1989, \apj, 346, 983
\bibitem[Elitzur (1992)]{eli92} Elitzur, Moshe 1992, Astronomical Masers (Astrophys.\ \& Space Sci.\ Library 170; Dordrecht: Kluwer)
\bibitem[Evans et al.\ (1970)]{eva70} Evans, N.\ J., II, Cheung, A.\ C., \& Sloanaker, R.\ M.\ 1970, \apjl, 159, L9
\bibitem[Evans et al.\ (1975)]{eva75} Evans, N.\ J., II, Morris, G., Sato, T., \& Zuckerman, B.\ 1975, \apj, 196, 433
\bibitem[Evans (1975)]{eva75b} Evans, N.\ J., II, 1975, \apj, 201, 112
\bibitem[Flower et al.\ (1990)]{flo90} Flower, D.\ R., Offer, A., \& Schilke, P.\ 1990, \mnras, 244, 4P
\bibitem[Forster et al.\ (1980)]{for80} Forster, J.\ R., Goss, W.\ M., Wilson, T.\ L., Downes, D., \& Dickel, H.\ R.\ 1980, \aap, 84, L1
\bibitem[Forster et al.\ (1982)]{for82} Forster, J.\ R., Graham, D., Goss, W.\ M., \& Booth, R.\ S.\ 1982, \mnras, 201, 7P
\bibitem[Forster et al.\ (1985)]{whi85} Forster, J.\ R., Goss, W.\ M., Gardner, F.\ F., \& Stewart, R.\ T.\ 1985, \mnras, 216, 35P
\bibitem[Garrison et al.\ (1975)]{gar75} Garrison, B.\ J., Lester, W.\ A., Jr., Miller, W.\ H., \& Green, S.\ 1975, \apjl, 200, L175
\bibitem[Gaume et al.\ (1991)]{gau91} Gaume, R.\ A., Johnston, K.\ J., Nguyen, H.\ A., Wilson, T.\ L., Dickel, H.\ R., Goss, W.\ M., \& Wright, M.\ C.\ H.\ 1991, \apj, 376, 608
\bibitem[Gaume et al.\ (1995)]{gau95} Gaume, R.\ A., Goss, W.\ M., Dickel, H.\ R., Wilson, T.\ L., \& Johnston, K.\ J.\ 1995, \apj, 438, 776
\bibitem[Green et al.\ (1978)]{gre78} Green, S., Garrison, B.\ J., Lester, W.\ A., Jr., \& Miller, W.\ H.\ 1978, \apjs, 37, 321
\bibitem[Guibert et al.\ (1978)]{gui78} Guibert, J., Elitzur, M., \& Nguyen-Q-Rieu 1978, \aap, 66, 395
\bibitem[Hill \& Hollenbach (1978)]{hil78} Hill, J.\ K.\ \& Hollenbach D.\ J.\ 1978, \apj, 225 390
\bibitem[Hoban et al.\ (1991)]{hob91} Hoban, S., Reuter, D.\ C., Mumma, M.\ J., \& Storrs, A.\ D.\ 1991, \apj, 370, 228
\bibitem[Hofner \& Churchwell (1996)]{hof96} Hofner, P.\ \& Churchwell, E.\ 1996, \aaps, 120, 283
\bibitem[Huettemeister et al.\ (1995)]{hut95} Huettemeister, S., Wilson, T.\ L., Mauersberger, R., Lemme, C., Dahmen, G., \& Henkel, C.\ 1995, \aap, 294, 667
\bibitem[Israel et al.\ (1973)]{isr73} Israel, F.\ P., Habing, H.\ J., \& de Jong, T.\ 1973, \aap, 27, 143
\bibitem[Jaffe \& Mart\'{\i}n-Pintado (1999)]{jaf99} Jaffe, D.\ T.\ \& Mart\'{\i}n-Pintado, J.\ 1999, \apj, 520, 162
\bibitem[Johnston et al.\ (1983)]{joh83} Johnston, K.\ J., Palmer, Patrick, Wilson, T.\ L., \& Bieging, J.\ H.\ 1983, \apjl, 271, L89
\bibitem[Kameya et al.\ (1990)]{kam90} Kameya, O., Morita, K.-I., Kawabe, R.\ \& Ishiguro, M.\ 1990, \apj, 355, 562
\bibitem[Kutner et al.\ (1971)]{kut71} Kutner, M., Thaddeus, P., Jefferts, K.\ B., Penzias, A.\ A., \& Wilson, R.\ W.\ 1971, \apjl, 164, L49
\bibitem[Kutner \& Thaddeus (1971)]{kt71} Kutner, M.\ \& Thaddeus, P.\ 1971, \apjl, 168, L67
\bibitem[Madden et al.\ (1986)]{mad86} Madden, S.\ C., Irvine, W.\ M., Matthews, H.\ E., Brown, R.\ D., \& Godfrey, P.\ D.\ 1986, \apj, 300, L79
\bibitem[Martin (1973)]{mar73} Martin, A.\ H.\ M.\ 1973, \mnras, 163, 141
\bibitem[Mart\'{\i}n-Pintado et al.\ (1985)]{mar85} Mart\'{\i}n-Pintado, J., Wilson, T.\ L., Henkel, C., \& Gardner, F.\ F.\ 1985, \aap, 142, 131
\bibitem[Mauersberger et al.\ (1987)]{mau87} Mauersberger, R., Henkel, C., \& Wilson, T.\ L.\ 1987, \aap, 173, 352
\bibitem[Mehringer et al.\ (1994)]{meh94} Mehringer, David M., Goss, W.\ M., \& Palmer, Patrick 1994, \apj, 434, 237
\bibitem[Mehringer et al.\ (1995)]{meh95} Mehringer, David M., Goss, W.\ M., \& Palmer, Patrick 1995, \apj, 452, 304
\bibitem[Minier et al.\ (1998)]{min98} Minier, V., Booth, R.\ S., \& Conway, J.\ E.\ 1998, \aap, 336, L5
\bibitem[Minier et al.\ (2000)]{min00} Minier, V., Booth, R.\ S., \& Conway, J.\ E.\ 2000, \aap, 362, 1093
\bibitem[Momose et al.\ (2001)]{mom01} Momose, M., Tamura, M., Kameya, O., Greaves, J.\ S., Chrysostomou, A., Hough, J.\ H., \& Morino, J.-I.\ 2001, \apj, 555, 855
\bibitem[Morisset et al.\ (2002)]{mor02} Morisset, C., Schaerer, D., Mart\'{\i}n-Hern\'{a}ndez, N.\ L., Peeters, E., Damour, F., Baluteau, J.-P., Cox, P., \& Roelfsema, P.\ 2002, \aap, 386, 558
\bibitem[Nerf (1972)]{ner72} Nerf, R.\ B., Jr.\ 1972, \apj, 174, 467
\bibitem[Osterbrock (1989)]{ost89} Osterbrock, D.\ E.\ 1989, Astrophysics of Gaseous Nebulae and Active Galactic Nuclei (University Science Books: Mill Valley)
\bibitem[Palmer et al.\ (1969)]{pal69} Palmer, Patrick, Zuckerman, B., Buhl, David, \& Snyder, Lewis E.\ 1969, \apjl, 156, L147
\bibitem[Palmer et al.\ (1984)]{pal84} Palmer, P., Gardner, F.\ F., \& Whiteoak, J.\ B.\ 1984, \mnras, 211, 41P
\bibitem[Pavlakis \& Kylafis (2000)]{pav00} Pavlakis, K.\ G.\ \& Kylafis, N.\ D.\ 2000, \apj, 534, 770
\bibitem[Pratap et al.\ (1992)]{pra92} Pratap, P., Snyder, L.\ E., \& Batrla, W.\ 1992, \apj, 387, 241
\bibitem[Pratap et al.\ (1994)]{pra94} Pratap, P., Menten, K.\ M., \& Snyder, L.\ E.\ 1994, \apj, 430, L129 (PMS94)
\bibitem[Pratap et al.\ (1999)]{pra99} Pratap, Preethi, Megeath, S.\ T., \& Bergin, Edwin A.\ 1999, \apj, 517, 799
\bibitem[Rots et al.\ (1981)]{rot81} Rots, A.\ H., Dickel, H.\ R., Forster, J.\ R., \& Goss, W.\ M.\ 1981, \apj, 245, L15
\bibitem[Sch\"{o}ier et al.\ (2002)]{sch02} Sch\"{o}ier, F.\ L., J{\o}rgensen, J.\ K., van Dishoeck, E.\ F., \& Blake, G.\ A.\ 2002, \aap, 390, 1001
\bibitem[Snyder et al.\ (1969)]{sny69} Snyder, L.\ E., Buhl, D., Zuckerman, B., \& Palmer, P.\ 1969, \prl, 22, 679
\bibitem[Sutton et al.\ (2001)]{sut01} Sutton, E.\ C., Sobolev, A.\ M., Ellingsen, S.\ P., Cragg, D.\ M., Mehringer, D.\ M., Ostrovskii, A.\ B., \& Godfrey, P.\ D.\ 2001, \apj, 554, 173
\bibitem[Thaddeus et al.\ (1971)]{tha71} Thaddeus, P., Wilson, R.\ W., Kutner, M., Penzias, A.\ A., \& Jefferts, K.\ B.\ 1971, \apjl, 168, L59
\bibitem[Thaddeus (1972)]{tha72} Thaddeus, P.\ 1972, \apj, 173, 317
\bibitem[Thompson (1999)]{tho99} Thompson, A.\ R.\ 1999, in Synthesis Imaging in Radio Astronomy II. Edited by G.\ B.\ Taylor, C.\ L.\ Carilli, and R.\ A.\ Perley. ASP Conference Series, 180, 11
\bibitem[Townes \& Cheung (1969)]{tow69} Townes, C.\ H.\ \& Cheung, A.\ C.\ 1969, \apjl, 157, L103
\bibitem[Townes \& Schawlow (1955)]{tow55} Townes, C.\ H.\ \& Schawlow, A.\ L.\ 1955, Microwave Spectroscopy; New York: McGraw-Hill
\bibitem[Tucker et al.\ (1971)]{tuc71} Tucker, K.\ D., Tomasevich, G.\ R., \& Thaddeus, P.\ 1971, \apj, 169, 429
\bibitem[Tucker et al.\ (1972)]{tuc72} Tucker, K.\ D., Tomasevich, G.\ R., \& Thaddeus, P.\ 1972, \apj, 174, 463
\bibitem[van der Tak et al.\ (2000)]{tak00} van der Tak, F.\ F.\ S., van Dishoeck, E.\ F., \& Caselli, P.\ 2000, \aap, 361, 327
\bibitem[Wadiak et al.\ (1988)]{wad88} Wadiak, E.\ J., Rood, R.\ T., \& Wilson, T.\ L.\ 1988, \apj, 324, 931
\bibitem[Walker (1999)]{wal99} Walker, R.\ C.\ 1999, in Synthesis Imaging in Radio Astronomy II. Edited by G.\ B.\ Taylor, C.\ L.\ Carilli, and R.\ A.\ Perley. ASP Conference Series, 180, 433
\bibitem[Walmsley \& Ungerechts (1983)]{wal83} Walmsley, C.\ M.\ \& Ungerechts, H.\ 1983, \aap, 122, 164
\bibitem[Walsh et al.\ (1998)]{wal98} Walsh, A.\ J., Burton, M.\ G., Hyland, A.\ R., \& Robinson, G.\ 1998, \mnras, 301, 640
\bibitem[Whiteoak \& Gardner (1983)]{whi83} Whiteoak, J.\ B.\ \& Gardner, F.\ F.\ 1983, \mnras, 205, 27P
\bibitem[Willner (1976)]{wil76} Willner, S.\ P.\ 1976, \apj, 206, 728
\bibitem[Wilson (1972)]{wil72} Wison, T.\ L.\ 1972, \aap, 19, 354
\bibitem[Wilson \& Henkel (1988)]{wil88} Wilson, T.\ L.\ \& Henkel, C.\ 1988, \aap, 206, L26
\bibitem[Wynn-Williams et al.\ (1974)]{wyn74} Wynn-Williams, C., Becklin, E., \& Neugebauer, G.\ 1974, \apj, 187, 473
\end{thebibliography}
\end{document}